\documentclass[aps,twocolumn,showpacs,amsmath,amssymb,prx,footinbib]{revtex4-1}

\usepackage{graphicx}
\usepackage{color}
\usepackage{amsmath}
\usepackage{psfrag}
\usepackage{natbib}
\usepackage{float}
\usepackage{verbatim}
\usepackage{hyperref} \hypersetup{colorlinks=true,linktoc=all,linkcolor=blue,breaklinks=true,citecolor=blue,urlcolor=blue}
\usepackage[normalem]{ulem}
\usepackage{lineno}
\usepackage{comment}

\definecolor{OliveGreen}{rgb}{0,0.6,0}

\begin{document}

\title{Readout of relaxation rates by nonadiabatic pumping spectroscopy}
\author{Roman-Pascal~Riwar$^{1,2}$, Beno\^it~Roche$^{1}$, Xavier~Jehl$^1$, Janine~Splettstoesser$^3$}
\affiliation{$^1$Univ. Grenoble Alpes, INAC SPSMS \& CEA INAC-SPSMS, F-38000, Grenoble, France\\
$^2$Peter Gr\"{u}nberg Institut (PGI-2) and JARA-Institute for Quantum Information, Forschungszentrum J\"{u}lich, D-52425 J\"{u}lich, Germany\\
$^{3}$Department of Microtechnology and Nanoscience, Chalmers University of Technology, S-41298 G\"{o}teborg, Sweden}

\date{\today}

\begin{abstract}
We put forward nonadiabatic charge pumping as a method for accessing the different charge relaxation rates as well as the relaxation rates of excited orbital states in double-quantum-dot setups, based on extremely size-limited quantum dots and dopant systems. 
The rates are obtained in a well-separated manner from \textit{plateaus}, occurring when comparing the steady-state current for reversed driving cycles. This yields a reliable readout independent of \textit{any} fitting parameters. 
Importantly, the nonadiabatic pumping spectroscopy essentially exploits the same driving scheme as the operation of these devices generally employs.
We provide a detailed analysis of the working principle of the readout scheme as well as of possible errors, thereby demonstrating its broad applicability. The precise knowledge of relaxation rates is highly relevant for the implementation of time-dependently operated devices, such as electron pumps for metrology or qubits in quantum information. 
\end{abstract}

\pacs{72.10.Bg, 73.23.Hk, 73.63.Kv, 06.20.Dk} 

\maketitle
\section{Introduction}
In order to render future nanoscale devices functional, coupling several elementary units to each other is essential. Double quantum dots (DQDs) represent a first step in this direction~\cite{vanderWiel02}: their orbital degrees of freedom allow for the implementation of solid-state qubits~\cite{Hayashi03,Petta05,Shi12,Koh12,Medford13} or qubit entanglement and readout protocols~\cite{Loss98,Elzerman03PRB,Koppens06}. Their precise tunability via separate external gates represents a possibility of controlled storage and transfer of single charges, e.g., in single-electron pumps~\cite{Buitelaar08a,Chorley12,Roche13,Pekola13,Connolly13}.  The dynamics of these devices is dictated by the lifetimes of excited states owing to internal transitions and relaxation times due to charge tunneling to external reservoirs. In particular, when these devices are operated at high frequencies, the knowledge of the different time scales is crucial. 

Previously, readout methods have usually been limited to a detection of one specific time scale of quantum dot devices. Substantial effort has for instance been made for the detection of spin relaxation rates, the inverse of the lifetime of excited spin states, using pulse-gating~\cite{Fujisawa01,Fujisawa02,Volk13}. 
Relaxation rates of charge are typically directly, albeit in general nontrivially, connected to the coupling strength to external reservoirs, such as electronic contacts or bosonic baths. For example, the charge relaxation rate of a single-level quantum dot is given by a product of the coupling strength to the leads and a factor which depends on temperature and the quantum state degeneracy.
The most easily accessible experimental data for the coupling strengths of multi-terminal quantum dots~\cite{Leturcq04} and DQDs~\cite{Stoof96,Nazarov93,Fujisawa98,Weber14} is obtained from a fit to the stationary-state current $I_\text{dc}$, which depends on all couplings to reservoirs, see App.~\ref{app_dc}. The drawback of this method is that the couplings cannot be read out independently of each other and several assumptions (concerning, e.g., their relative magnitude or the impact of degeneracies on the relaxation rates) have to be made to extract each rate from an appropriate fitting procedure. This is often possible for relatively large quantum dots obtained by means of lithographic methods, but it becomes an issue for ultra-small systems such as dopants, where the coupling strengths remain very hard to engineer. 

Instead, in order to directly measure the relaxation rates (the inverse of the time scales of the charge dynamics), time-resolved measurements have been necessary~\cite{Feve07,Mahe10,Beckel14}. They possibly even yield the full counting statistics, when a nearby quantum point contact provides the necessary sensitivity for single-electron detection~\cite{Gustavsson06,Fricke07,Fujisawa06}. An alternative for such approaches exploits the finite-frequency noise, which contains information on dwell times~\cite{Parmentier12}, requiring the challenging measurement of current-current correlations. Finally, radio-frequency reflectometry has recently been used  as a tool to study tunnel couplings in dopant-based systems~\cite{House15}.

In this paper, we propose a fitting-free readout scheme, making use of a detection of steady-state currents resulting from nonadiabatic charge pumping. It enables at the same time to read out two distinct classes of relaxation rates of DQDs from the same method and device. These are (1) the charge relaxation rates due to tunneling into electronic contacts, with their dependence on the ground state degeneracy of the different dots, and (2) the relaxation rate of excited orbital states of the DQD resulting from inelastic effects. Complications arising in the study of time-resolved quantities or correlation functions, as well as difficulties related to backaction effects, as they are  induced by other types of detectors~\cite{Schulenborg14,Gustavsson06,Fricke07}, are circumvented.  Nevertheless, the read-out time is not significantly enhanced with respect to the standard dc readout scheme, see App.~\ref{app_dc}. Since charge pumping results from the time-dependent modulation of local quantum dot gates, the \textit{detection method} proposed here relies on an \textit{operation principle} of DQDs that has been well established in recent years in the context of pumping and qubit operation. In particular, no complicated shaping of the driving signals is required -- simple harmonic driving is sufficient.

The periodic modulation of the local gates applied to a DQD allows for quantized charge transfer, which is, for instance, desirable for the implementation of a quantum standard for the current~\cite{Pothier92,Chorley12,Connolly13,Pekola13}. Therefore the modulation needs to be adiabatic; whenever the driving is fast with respect to different relaxation processes of the DQD system, errors occur~\cite{Keller96,Roche13}. These errors are at the basis of the implementation of our detection scheme. It uses the time-averaged current through the DQD,  $I=\int_0^\tau  I(t) \ dt/\tau$, due to a driving cycle of the gates, with cycle period $\tau$, and compares it to the time-averaged current for the reversed pumping cycle, denoted by $I'$. While in the adiabatic limit $I'=-I$, for nonadiabatic driving the reversed current $I'$ is nontrivially different from $I$. Depending on the different ranges of working points of the pump, extended regions in the stability diagram can be pinpointed, in which the ratio between these two currents yields the value
\begin{equation}\label{eq_current_ratio}
\frac{I}{I'}=e^{\gamma_x \delta t_x}
\end{equation}
for different relaxation rates  $\gamma_x$. These regions are shown in Fig.~\ref{fig_model}~(d), indicating that each rate can be read out \textit{independently} of the other rates from \textit{plateaus}. 
The relation between currents of reversed cycles, Eq.~(\ref{eq_current_ratio}), is the main ingredient to the proposed detection scheme. It thereby constitutes a differential measurement: the outcomes of two current measurements are compared in a way that all other dependencies on free parameters are cancelled. The relaxation rates can hence be read out \textit{without any further fitting procedure}. Importantly, no conditions on the relative asymmetry of the rates is required.  The proposed scheme is therefore expected to be highly advantageous for future experimental applications.

\section{Double quantum dot} 

\subsection{Model}

\begin{figure}[b]
\includegraphics[width=3.4in]{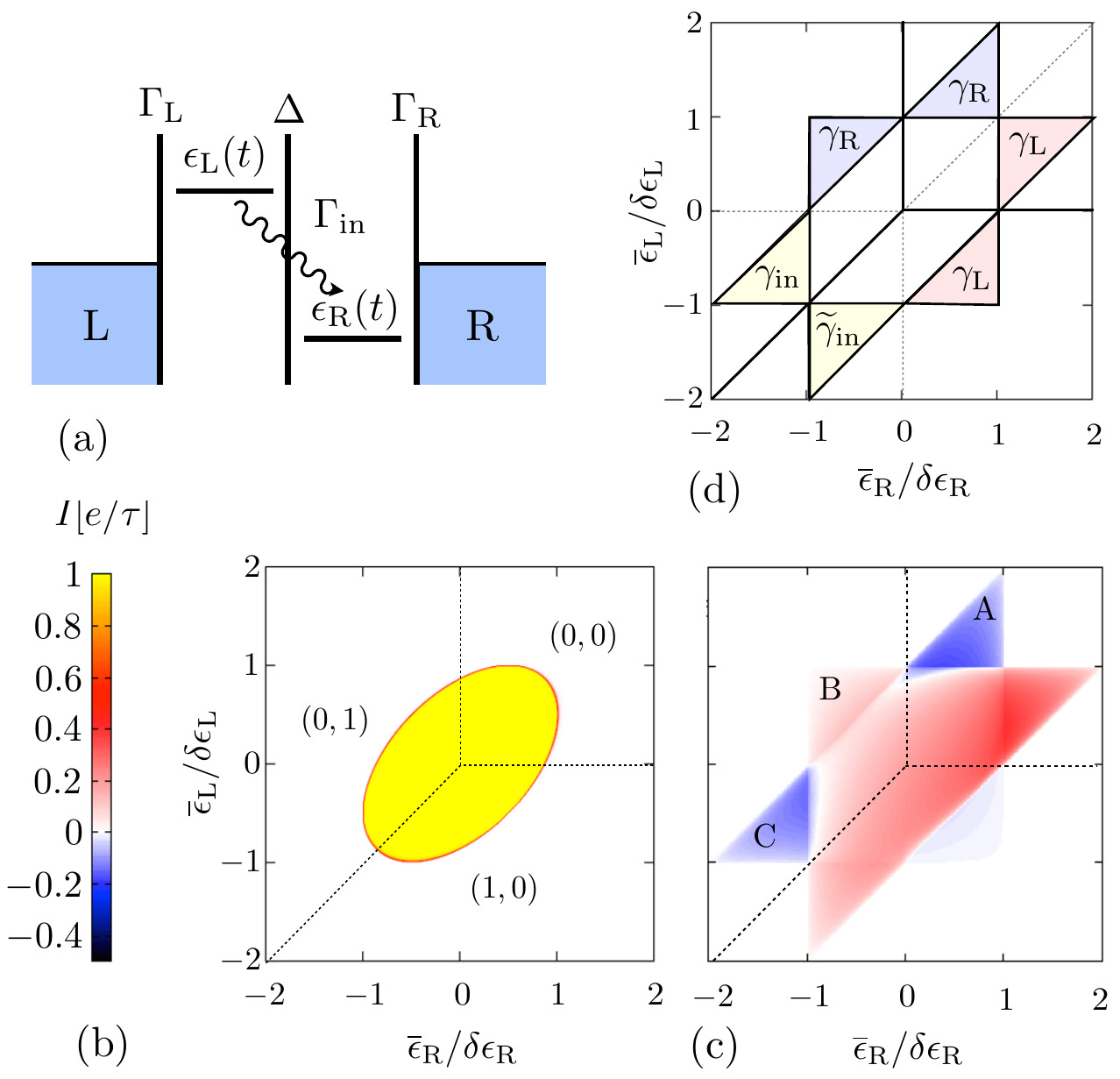}
\caption{(a) Energy landscape of the DQD. (b) Stability diagram of the DQD as a function of  $\bar{\epsilon}_\mathrm{L}$ and $\bar{\epsilon}_\mathrm{R}$, with the time-averaged quantized current in the adiabatic regime. We take $\Omega/(2\pi\Gamma)=10^{-3}$, $\Gamma_\text{L}=\Gamma_\text{R}$, $\delta\epsilon_\text{L,R}/\Gamma=10^{4}$, $\Omega/(2\pi\Gamma_\text{in})=1/2$, and $\varphi=\pi/3$. (c) Time-averaged current in the nonadiabatic regime $\Omega/(2\pi\Gamma)=10^{-1}$, with the other parameters as in (b). The asymmetry of the displayed signal stems from the driving direction. (d) Overview of  plateau regions in which the detection scheme provides the charge relaxation rates ($\gamma_\text{R}$ and $\gamma_\text{L}$) and the relaxation rate of the excited orbital state ($\gamma_\text{in},\tilde{\gamma}_\text{in}$).
\label{fig_model}}
\end{figure}

We consider a DQD consisting of two serially coupled single-level dots, as depicted in Fig.~\ref{fig_model}~(a). We assume that the Coulomb interaction in the DQD is the largest energy scale. This means that the DQD can be in the states $\{|0\rangle,|\text{L}\rangle,|\text{R}\rangle\}$, with either no extra charge on the DQD, or with one electron (with spin $\uparrow,\downarrow$) on the left or right dot. With this choice we focus on one of the regions of the DQD's stability diagram; equivalent considerations can be done for any other region of higher occupation numbers. The Hamiltonian of the isolated DQD is 
\begin{equation}\label{eq_H_DQD}
H(t)=\epsilon_\text{L}|\text{L}\rangle\langle\text{L}|+\epsilon_\text{R}|\text{R}\rangle\langle\text{R}|+\frac{\Delta}{2}\left(|\text{L}\rangle\langle\text{R}|+|\text{R}\rangle\langle\text{L}|\right). 
\end{equation}
The single-particle energies on each dot depend on time, $\epsilon_\text{L}(t)=\overline{\epsilon}_\text{L}+\delta\epsilon_\text{L} \sin(\Omega t)$ and $\epsilon_\text{R}(t)=\overline{\epsilon}_\text{R}+\delta\epsilon_\text{R} \sin(\Omega t+\varphi)$ with the driving frequency $\Omega=2\pi/\tau$ and the phase-shift $\varphi$. For our detection scheme -- the same regime in which typical quantized charge pumps are realized~\cite{Pothier92,Chorley12,Roche13,Pekola13,Connolly13} -- we require the driving amplitudes to be large with respect to the temperature broadening, $\delta\epsilon_\alpha\gg k_\mathrm{B}T$ for $\alpha=\text{L,R}$. Electron tunneling between the two dots takes place with the tunneling amplitude $\Delta$. In the absence of driving,  as indicated in the stability diagram in Fig.~\ref{fig_model}~(b), the system can be found in different stable charge configurations $(n_\text{L},n_\text{R})$, where $n_\text{L/R}$ is the occupation number of the left/right dot.
The left and right dots are each tunnel-coupled to one (noninteracting) electronic reservoir, kept at the same electrochemical potential $\mu$, with which they can exchange charge. Inelastic transitions between the dot levels (for instance, mediated through an electron-phonon coupling~\cite{Fujisawa98,Hayashi03,Hu05}) are modelled by coupling of the dots to a common bosonic reservoir. The coupling to these reservoirs is characterized by the coupling strengths $\Gamma_\text{L}$ and $\Gamma_\text{R}$  with $\Gamma=\sum_{\alpha=\text{L,R}}\Gamma_\alpha$ for the electronic reservoirs, and $\Gamma_\text{in}$ for the bosonic reservoir. Refering to typical experiments, the coupling between the dots and between the dots and the reservoirs is assumed to be weak, $\Delta, \Gamma_\text{in},\Gamma_\alpha\ll k_\text{B}T$~\cite{Roche13}. We set $\hbar=1$.

\subsection{Dynamics and relaxation rates}\label{sec_dynamics_rates}
We describe the dynamics of the DQD based on its reduced density matrix, $P(t)$. It has both diagonal and nondiagonal parts which we represent in a vector $P=(P_\text{d},P_\text{nd})$. The diagonal part contains the occupation probabilities of the dot states $P_\text{d}=(p_0,p_\text{L},p_\text{R})$ and the non-diagonal part $P_\text{nd}=(p_\text{LR},p_\text{RL})$ the coherences. The time evolution of the reduced density matrix  is given by the Master equation~\footnote{The time-convolution-less dynamics due to the reservoirs is a valid approximation, if the correlations in the reservoirs decay much faster than the time scale set by the system-reservoir coupling, here requiring the experimental relevant regime $\Gamma\ll k_\text{B}T$.}
\begin{equation}
\label{eq_master}
\dot{P}(t)=-i\left[H(t),P(t)\right]+W(t)P(t)\ ,
\end{equation}
see for example Refs.~\cite{Grifoni98,Wunsch05,Riwar10}. 
The system has a coherent internal Hamiltonian dynamics, described by  $H(t)$ of the isolated DQD, Eq.~(\ref{eq_H_DQD}), while the kernel $W(t)$ describes the dissipative coupling to external reservoirs. The time-dependent driving of the DQD gives rise to an explicit time dependence of $H(t)$ and $W(t)$. In the regime of weak coupling, $\Delta,\Gamma_\alpha,\Gamma_\text{in}\ll k_\mathrm{B} T$, we obtain the kernel $W$ from Fermi's golden rule, where only transitions between diagonal elements of the reduced density matrix matter~\footnote{There are two reasons for that: (1) $W$ does not couple the diagonal and nondiagonal sector due to $\Delta,\Gamma\ll k_\text{B}T$. (2) We assume that the coherences of $P(t)$ that can arise at the level crossings decay very rapidly and are therefore not relevant for the \textit{dissipative} dynamics of the system.}. There are two contributions $W_\text{dd}(t)=\left[W_\text{tunn}(t)\right]_\text{dd}+\left[W_\text{in}(t)\right]_\text{dd}$. The first stems from tunneling to the electronic reservoirs,
\begin{eqnarray}\label{eq_Wtun}
&&\left[W_\text{tun}(t)\right]_\text{dd}=\\
&&\left(\begin{array}{ccc}
-2\Gamma_{\text{L}}f_{\text{L}} & \Gamma_{\text{L}}f_{\text{L}}^{-} &0\\
2\Gamma_{\text{L}}f_{\text{L}} & -\Gamma_{\text{L}}f_{\text{L}}^{-} & 0\\
0 & 0 &0\end{array}\right)+\left(\begin{array}{ccc}
-2\Gamma_{\text{R}}f_{\text{R}} & 0 & \Gamma_{\text{R}}f_{\text{R}}^{-}\\
0 & 0& 0\\
2\Gamma_{\text{R}}f_{\text{R}} & 0 & -\Gamma_{\text{R}}f_{\text{R}}^{-}\end{array}\right)\nonumber
\end{eqnarray}
with the Fermi function $f_\alpha=1/\left(1+e^{\epsilon_\alpha(t)/k_\text{B}T}\right)$ (and $f_\alpha^-=1-f_\alpha$) for left and right reservoirs, $\alpha=\text{L,R}$. Because of the large amplitudes of the driving parameters
the Fermi functions can be approximated by Heaviside-functions, $f_\alpha\approx\theta(-\epsilon_\alpha)$. It furthermore turns out that during a driving cycle, the dynamics due to tunneling to the reservoirs is always separately determined by either of the two kernel contributions given in Eq.~(\ref{eq_Wtun}), see also Sec.~\ref{sec_pumping_current} and App.~\ref{app_triangleA}. This means that the charging and discharging of the DQD always takes place either between the left dot and the left reservoir or between the right dot and the right reservoir at a time. The dynamics of the DQD due to charging and discharging is therefore governed by the \textit{charge relaxation rates} $\gamma_\text{L}=\Gamma_\text{L}(1+f_\text{L}(\epsilon_\text{L}))$ and $\gamma_\text{R}=\Gamma_\text{R}(1+f_\text{R}(\epsilon_\text{R}))$, which are obtained as the absolute value of the non-zero eigenvalues of the two distinct kernel contributions of Eq.~(\ref{eq_Wtun}), see Refs.~\cite{Splettstoesser10,Schulenborg15}.

The kernel contribution for inelastic transitions between dot levels is
\begin{equation}
\left[W_\text{in}(t)\right]_\text{dd}=\left(\begin{array}{ccc}
0 & 0 & 0\\
0 & -\gamma_\text{in} & \tilde{\gamma}_\text{in}\\
0 & \gamma_\text{in} & -\tilde{\gamma}_\text{in}\end{array}\right),
\end{equation}
with the relaxation rates $\gamma_\text{in}$ ($\tilde{\gamma}_\text{in}$) from the right to the left (left to right) dot level.  These rates can in general depend nontrivially on the energy detuning $\epsilon_\text{L}-\epsilon_\text{R}$ of the dot levels. For very small systems, they are however well approximated as $\gamma_\text{in}=\Gamma_\text{in}\theta(\epsilon_\text{L}-\epsilon_\text{R})$ ($\tilde{\gamma}_\text{in}=\Gamma_\text{in}\theta(\epsilon_\text{R}-\epsilon_\text{L})$) for a large level detuning $|\epsilon_\text{L}-\epsilon_\text{R}|\gg k_\text{B}T$~\cite{Roche13}. The large amplitudes of the driving indeed ensure that the time interval around the crossing, in which $|\epsilon_\text{L}-\epsilon_\text{R}|\leq k_\text{B}T$, is so short that during this time no inelastic processes take place. Then, $\Gamma_\text{in}$ is energy-independent on the scale of the distance between triple points in the stability diagram. Unless stated otherwise, this is our default assumption. In addition, we show in Sec.~\ref{sec_inelastic_E} how to readout a generally energy dependent inelastic relaxation rate from the same readout scheme.

\subsection{Pumping current}\label{sec_pumping_current}
The quantity of interest for the rate readout is the time-averaged charge current across the DQD, $I=\int_0^\tau I(t) \ dt/\tau $ due to periodic driving of the gates. The time-resolved current is defined as $I(t)=\big(\dot{N}_\text{L}-\dot{N}_\text{R}\big)/2$, where $\dot{N}_\alpha$ is the time derivative of the electron number in reservoir $\alpha$, and computed as $I(t)=\text{tr}\left[W_I(t)P(t)\right]$. Here, $W_I$ is derived from $W$ by retaining all elements that change the charge of the system, and adding an appropriate sign, as to whether the charge is leaving/entering the DQD to the left/right reservoir, see App.~\ref{app_triangleA}.

Periodic driving by applying time-dependent gates means that closed trajectories in the stability diagram are performed. Thereby the energy levels of the two dots can cross and change their relative position, or the two levels can be moved above and below the common Fermi energy of the leads. [Two examples are shown in the upper panels of Figs.~\ref{fig_trajectories} and \ref{fig_numeric}.] This can eventually lead to hopping of electrons between the two dots, and between the DQD and the reservoirs, depending on the working point of the pump, defined by the pair $(\bar{\epsilon}_\text{L},\bar{\epsilon}_\text{R})$, and on the velocity of the two modulated levels along the driving cycle, $\dot{\epsilon}_\text{L/R}$.
When driven adiabatically, namely, when the system can follow the driving, either exactly zero or exactly one electron are transferred through the DQD per pumping cycle, depending on the working point, see Fig.~\ref{fig_model}~(b).~\footnote{Except for smeared regions in between the two regimes, which occur due to inevitable, nonadiabatic effects at driving cycles that traverse the vicinity of a triple point.} If only the working point is chosen in a way that a triple point is included in the parameter trajectory, there are two different times at which the energy levels cross during one cycle, once above and once below the Fermi energy. Then, exactly one electron is passed from one reservoir to the other. In this regime of \textit{quantized pumping} the dc current  is $I_{\text{ad}}=\pm e/\tau$, where the sign depends on the direction of the driving cycle. For all other working points, we have $I_{\text{ad}}=0$.

The crossover from the adiabatic to the nonadiabatic regime can be reached by a relatively small increase of the driving frequency, if the amplitudes of the driving are sufficiently large. Recently, a similar crossover from adiabatic, quantized electron pumping to nonadiabatic pumping has been shown for an ultrasmall DQD realized with single atomic dopants~\cite{Roche13}. There, the driving amplitudes are large, $\delta\epsilon_\alpha\gg k_\text{B}T$, leading to a large driving \textit{speed}, $\dot{\epsilon}_\text{L/R}$. 
Then, the dc current can generally be separated into $I=I_{\text{ad}}+I_{\text{non}}$, where $I_{\text{ad}}$ is the adiabatic, quantized current and $I_{\text{non}}$ is the purely nonadiabatic current contribution.
The nonadiabatic driving strongly modifies the pumping current, leading to less than one electron pumped per cycle in the central region and to triangular shaped structures with finite charge current, in regions where $I_{\text{ad}}=0$, see Fig.~\ref{fig_model}~(c). 
More specifically, these nonadiabatic effects emerge due to two different conditions and are related to two different types of (missed) processes~\cite{Roche13}: (1) When frequency and amplitude are large, namely $\Omega\delta\epsilon\sim\Delta^2$, coherent Landau-Zener transitions between ground and excited state occur. Here, $\Omega\delta\epsilon$ stands for the order of magnitude of the detuning speed. In the setup considered here, this means that when the two orbital levels cross, the electron does possibly not hop to the lower-lying energy level at the moment of the crossing. The probability for an electron occupying the excited state after a level crossing is given by $p_\text{LZ}=1-\exp{(-\pi\Delta^2/2|\dot{\epsilon}_\text{L}-\dot{\epsilon}_\text{R}|)}$. (2) When the frequency is as large~\footnote{More precisely, the time intervals between different charging events defined in Figs.~\ref{fig_trajectories}
and~\ref{fig_numeric} should satisfy $\Delta\tau_\text{c}^{-1}$, $\Delta t^{-1}$, $\Delta\tau^{-1}$, $\Delta t'^{-1}$, $\Delta \tilde{\tau}_\text{c}^{-1}$, $\Delta \tilde{t}^{-1}$, $\Delta\tilde{\tau}^{-1}$, $\Delta \tilde{t}'^{-1}\ll\Gamma$. For most working points, these time intervals scale as $\Omega^{-1}$, except for working points where two charging events become very close. This is the case, e.g., at the red rim of the circle of the quantised pumping current, see Fig.~\ref{fig_model}~(b), where adiabaticity is locally broken even though $\Omega\ll\Gamma$.}
as $\Omega\sim\Gamma$, the charge relaxation rates of the two dots are of the order of the inverse time scale of the driving. Then, the dissipative relaxation of the system is possibly not complete between two crossings of the dot levels or of one of the dot levels with the Fermi energy of the reservoirs. The resulting $I_{\text{non}}$ is nontrivially sensitive to the pumping direction.
In fact, we show in the following that for the working points in the triangular regions, where $I_{\text{ad}}=0$ and $I_{\text{non}}\neq0$, see Fig.~\ref{fig_model}~(c), the ratio of the pumping current for a forward and backward driving cycle, $I/I'=I_{\text{non}}/I_{\text{non}}'$, provides a direct signature of the relaxation rates of the system.

In the regime of interest, $\delta\epsilon_\alpha\gg k_\text{B}T,\Delta$, the coherent and the incoherent, dissipative dynamics of the DQD connected to external reservoirs can be separated. The time interval during which a single crossing of the two DQD levels takes place, $|\epsilon_\text{L}(t)-\epsilon_\text{R}(t)|<\Delta$, can be given as $\delta t_\text{cross}\sim\Delta/|\dot{\epsilon}_\text{L}-\dot{\epsilon}_\text{R}|$. When the condition  $\delta t_\text{cross}\ll\Gamma^{-1}$ is fulfilled, as ensured by the large driving amplitudes,  the dynamics during this interval is given only by the coherent part of the master equation, Eq.~(\ref{eq_master}), $\dot{P}=-i[H,P]$. Since this crossing time interval is much shorter than any other relevant time scales, such as the pumping period $\tau$, one can neglect its duration, and one can treat it like the Landau-Zener problem discussed above. In contrast, in all time intervals between any two such crossings, the level separation is large, and thus the time-evolution of $P$ is determined by dissipative dynamics described by the kernel $W$ only.

\section{Readout of charge relaxation rates}\label{sec_charge}
\begin{figure}
\includegraphics[width=0.33\textwidth]{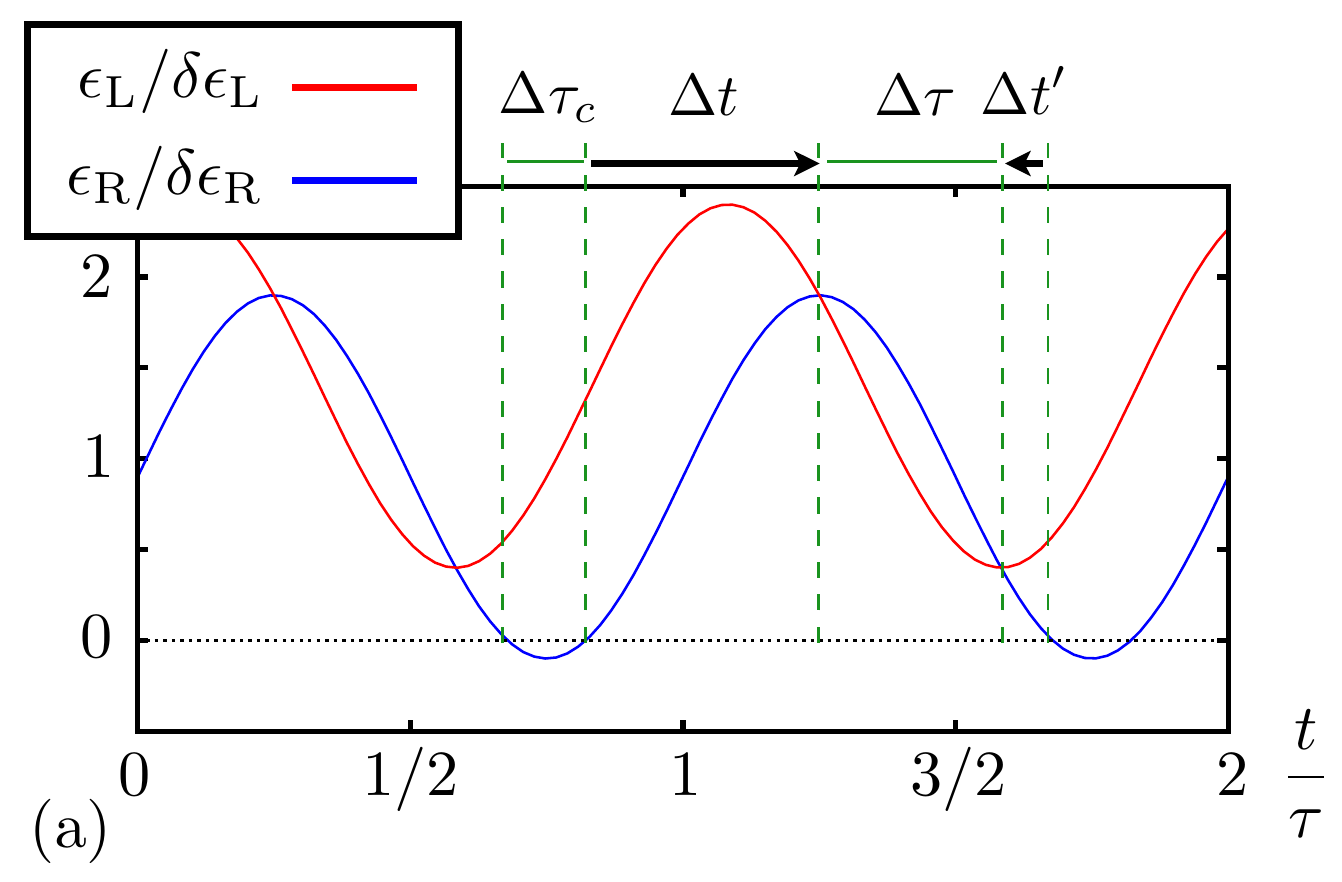}
\includegraphics[width=0.48\textwidth]{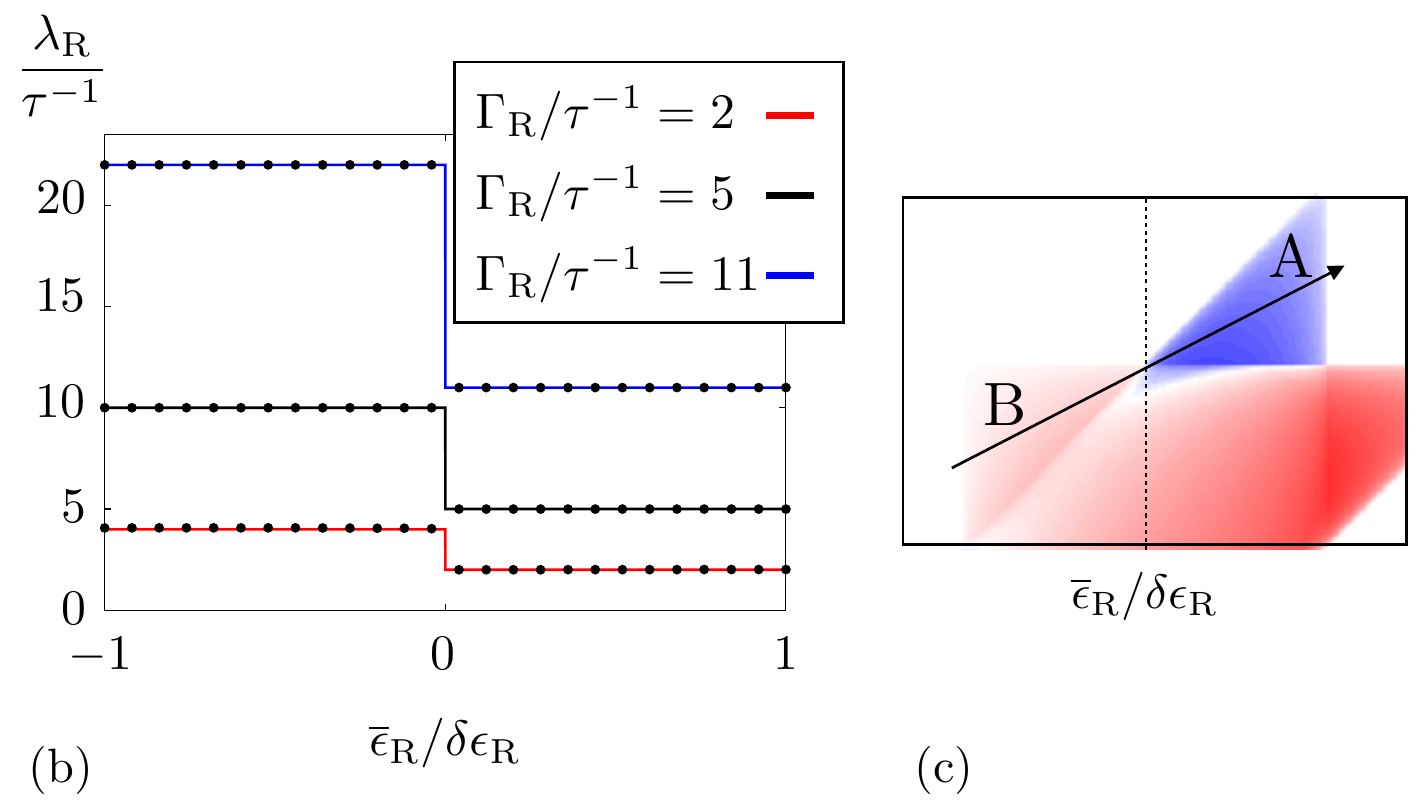}
\caption{(a) Example of $\epsilon_\text{L,R}$ as a function of time, in units of the modulation amplitude $\delta\epsilon_\text{L,R}$, showing the generic time evolution of the left and right energy levels in triangle A. (b) Extracted charge relaxation rate $\gamma_\text{R}^\text{ex}$ (dots), and actual charge relaxation rate $\gamma_\text{R}$ (lines) for different values of $\Gamma_\text{R}$ along the arrow in the working point space indicated in (c) showing a zoom into the stability regions of Fig.~\ref{fig_model}~(c). The arrow can be parametrized as $(\overline{\epsilon}_\text{L},\overline{\epsilon}_\text{R})=(\delta\epsilon,0)+(1,2)s$ with $s\in[-\delta\epsilon/2,\delta\epsilon/2]$ and $\delta\epsilon_\text{L}=\delta\epsilon_\text{R}=10^{4}\Gamma$. In (b) $\Gamma_\text{L}=5\tau^{-1}$, whereas the remaining parameters are as in Fig.~\ref{fig_model}~(c).}
\label{fig_trajectories}
\end{figure}
 We first focus on the readout of the charge relaxation rates from both the left and the right dot due to tunneling of electrons to the respective reservoir, $\gamma_\text{L}$ and $\gamma_\text{R}$, see Sec.~\ref{sec_dynamics_rates}. The value of the charge relaxation rate differs, depending on whether the left or the right dot is involved in the relaxation process and on whether the observed relaxation leads to the charging or the discharging of the respective dot. The reason for the latter difference is the spin-degeneracy of the dots' single levels, which doubles the number of accessible states for the charging process with respect to the discharging~\cite{Splettstoesser10,Reckermann10,Beckel14}. The resulting four different charge relaxation rates -- $\gamma_\text{L}$ and $\gamma_\text{R}$ both for charging and discharging -- can be read out from the triangular regions indicated in Fig.~\ref{fig_model}~(d).  The measurement of the charge relaxation rates of the right dot, $\gamma_\text{R}$, is possible at the working points within triangle A, where the \textit{discharging} of the right dot is found to take place with the rate $\gamma_\text{R}=\Gamma_\text{R}\left(1+f(\epsilon_\text{R}\right))\approx\Gamma_\text{R}$, and at triangle B, where the \textit{charging} of the right dot is extracted to take place with the rate $\gamma_\text{R}=\Gamma_\text{R}\left(1+f(\epsilon_\text{R}\right))\approx2\Gamma_\text{R}$). The  charge relaxation rate $\gamma_\text{L}$ can be observed in the triangles on the opposite side, see Fig.~\ref{fig_model}~(d).

 For triangle A, the two level crossings occur above the Fermi energy, see Fig.~\ref{fig_trajectories}~(a), where both quantum dots would be empty in the adiabatic limit. However, if the driving is too fast for the right dot to discharge, the pumping current across the DQD is finite. To bring out this feature, let us consider the case $\Gamma_\text{L}\gg\Gamma_\text{R}$, where tunneling between the left dot and the left reservoir occurs almost instantaneously, see App.~\ref{app_triangleA}. Note that this limit is chosen only for simplicity of the analytic treatment, and the readout scheme remains valid for arbitrary tunnel coupling ratios, see Fig.~\ref{fig_trajectories}~(b) and Fig.~\ref{fig_precision}. Then, the result for the time-averaged current of triangle A can be written in the intuitive form
\begin{equation}\label{eq_I_A}
\frac{\overline{I}_\text{A}}{e/\tau}=-n^\text{R}e^{-\Gamma_\text{R}\Delta t}\left[i_\text{LZ,1}+i_\text{LZ,2}+i_\text{in}\right]\ .
\end{equation}
Here, $n^\text{R}=\left(1-e^{-2\Gamma_\text{R}\Delta\tau_\text{c}}\right)/\left(1-e^{-\Gamma_\text{R}(\Delta\tau_\text{c}+\tau)-\Gamma_\text{in}\Delta\tau}p_\text{LZ}^2\right)$ is the occupation of the right dot, before $\epsilon_\text{R}(t)$ crosses the Fermi energy from below. [See Fig.~\ref{fig_trajectories}~(a) for the definition of the time intervals, $\Delta t,\Delta t', \Delta \tau$ and $\Delta \tau_c$.] When the subsequent discharging of the right dot is not complete before the crossing of the two dot levels, $e^{-\Gamma_\text{R}\Delta t}\neq0$, a finite current can arise due to a transfer of charge from the right to the left dot. At the first crossing of the dot levels, at the end of the time interval $\Delta t$, this happens with probability $i_\text{LZ,1}=1-p_\text{LZ}$. Otherwise, if the electron stays on the dot during the interval $\Delta \tau$ until the second level crossing occurs at the end of this time interval, again a transfer of charge can take place; this happens with the amplitude $i_\text{LZ,2}=e^{-(\Gamma_\text{R}+\Gamma_\text{in})\Delta\tau}p_\text{LZ}(1-p_\text{LZ})$. However, with the amplitude  $i_\text{in}=\frac{\Gamma_\text{in}}{\Gamma_\text{R}+\Gamma_\text{in}}(1-e^{-(\Gamma_\text{R}+\Gamma_\text{in})\Delta\tau})p_\text{LZ}$ the electron can already be transferred to the left dot during the time interval $\Delta\tau$ due to an inelastic relaxation from the higher to the lower-lying dot level.

When reversing the pumping cycle, the coherent Landau-Zener processes occurring at the crossings remain the same, and the only relevant change concerns the time before the first crossing between the two levels at the end of the time interval $\Delta t$. This means that in order to obtain the current for a reversed pumping cycle at triangle A, $\overline{I}_\text{A}'$, we need to replace $\Delta t\rightarrow\Delta t'$. In Eq.~\eqref{eq_I_A}, it is only the discharging probability $e^{-\Gamma_\text{R}\Delta t}$ that depends on this time interval, and consequently the ratio of currents provides
\begin{equation}\label{eq_ratio_GammaR}
\frac{\overline{I}_\text{A}}{\overline{I}_\text{A}'}=e^{-\Gamma_\text{R}(\Delta t-\Delta t')}\ .
\end{equation}
Importantly, while $\Delta t$ and $\Delta t'$ each depend on the working point $\left(\overline{\epsilon}_\text{L},\overline{\epsilon}_\text{R}\right)$, their difference $\Delta t-\Delta t'$ does not within the area A, see App.~\ref{app_interval}. Therefore, the ratio $\overline{I}_\text{A}/\overline{I}_\text{A}'$ gives rise to a plateau value within this triangle, enabling a reliable readout. Note, that the relaxation rate $\gamma_\text{R}=\Gamma_\text{R}$ (discharging) is here the only free parameter, since $\Delta t-\Delta t'$ is determined by the external driving parameters. Similarly, the pumping current at triangle B allows for the readout of the charging rate, $\gamma_\text{R}\approx2\Gamma_\text{R}$.

In Fig.~\ref{fig_trajectories}~(b) we show the charge relaxation rate $\gamma_\text{R}$ for different values of the coupling strength $\Gamma_\text{R}$ across a straight line passing through the triangles A and B, in Fig.~\ref{fig_model}~(c). The stepwise constant function reveals the announced factor of 2 between the charging and the discharging rate due to spin degeneracy.  In this plot, we show both the actual charge relaxation rate $\gamma_\text{R}$ (full lines) as well as the extracted rate from a possible detection, $\gamma_\text{R}^\text{ex}=\ln\left(I/I'\right)/\left(\Delta t'-\Delta t\right)$, where no approximations concerning the asymmetry between the coupling strengths $\Gamma_\text{L}$ and $\Gamma_\text{R}$ were made (dots).
 Differences between the full and the dotted lines are too small to be perceived from this plot. We show a more detailed discussion of the readout precision in Sec.~\ref{sec_precision}. 
 
\section{Readout of the relaxation rate of the excited state}\label{sec_inelastic}
Intriguingly, the detection scheme proposed here allows one to read out not only the charge relaxation rates of the two dots, but also  the internal inelastic relaxation rate of the excited orbital state of the DQD given by  $\gamma_\text{in}$ and $\widetilde{\gamma}_\text{in}$ respectively, which are determined by the energy-independent $\Gamma_\text{in}$. This readout can be done at working points within the lower left triangles in Fig.~\ref{fig_model}~(d).

\begin{figure}
\includegraphics[width=0.42\textwidth]{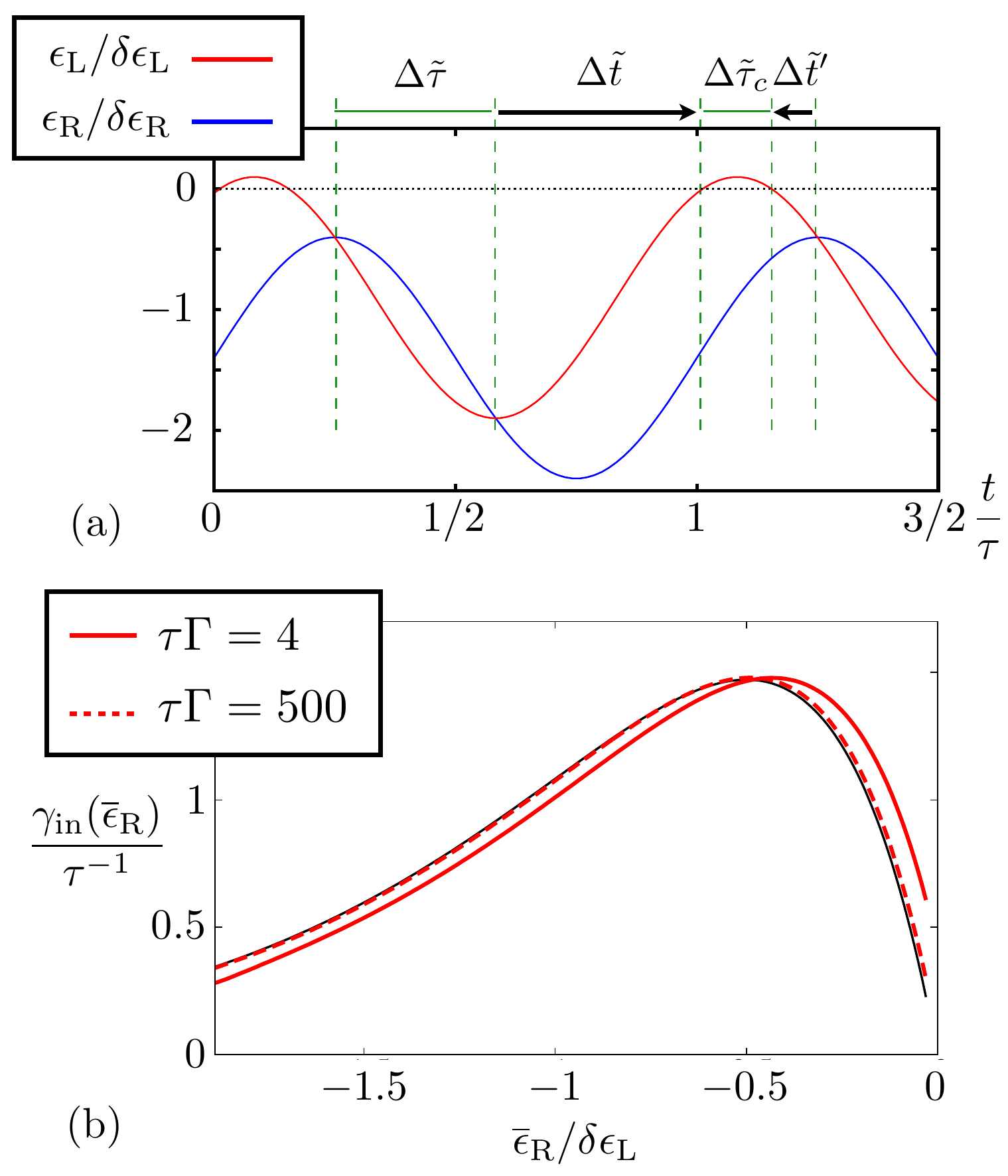}
\caption{(a) Example of $\epsilon_\text{L,R}$ as a function of time, in units of the modulation amplitude $\delta\epsilon_\text{L,R}$, showing the generic time evolution of the left and right energy levels in triangle C. (b) Extracted inelastic rate $\gamma_\text{in}^\text{ex}$ (dashed), based on the current signal ratio $\overline{I}/\overline{I}'$  (see Eq.~\eqref{eq_inel_energy_dep}) for different values of $\Gamma$, and actual rate $\gamma_\text{in}$ (solid) as a function of $\overline{\epsilon}_\text{R}$. The readout is performed along the line in working point space which is parametrized by $(\overline{\epsilon}_\text{L},\overline{\epsilon}_\text{R})=(-0.9\delta\epsilon_\text{L},0)+(0,1)s$, with $s=[-1.9\delta\epsilon_\text{L},0]$. The remaining parameters are $\varphi=\pi/2$, $\Gamma_\text{in}=2\tau^{-1}$, $\omega_\text{co}=0.5\delta\epsilon_\text{L}$, $\Gamma_\text{L}=\Gamma_\text{R}$, $\delta\epsilon_\text{R}/\delta\epsilon_\text{L}=0.025$ and $\delta\epsilon_\text{L}/\Gamma=10^4$.}\label{fig_numeric}
\end{figure}

For example, for working points within triangle C of Fig.~\ref{fig_model}~(c) both level crossings occur below the Fermi energy, and only the energy level of the left dot exceeds the Fermi energy for a finite discharging time interval, $\Delta\tilde{\tau}_c$, see Fig.~\ref{fig_numeric}~(a). In triangle C, Landau-Zener transitions and inelastic relaxation between the levels dominate and nonadiabatic effects due to the coupling to the electron reservoirs are unimportant. For simplicity, we therefore assume the tunneling dynamics to be faster than the driving, $\Gamma_\text{L},\Gamma_\text{R}\gg\Omega$. In this limit, the current within triangle C is given by the simple analytic expression (see App.~\ref{app_triangleC1}),
\begin{equation}\label{eq_current_triangle_C_const}
\frac{\overline{I}_\text{C}}{e/\tau}=-n_\text{c}^\text{L}e^{-\Gamma_\text{in}\Delta\tilde{t}}\ .
\end{equation}
The current is nonzero only if the left dot occupation immediately after the two crossings (namely at the end of the interval $\Delta\tilde{\tau}$), $n_\text{c}^\text{L}=(1+e^{-\Gamma_\text{in}\Delta\widetilde{\tau}}[1-2p_\text{LZ}])p_\text{LZ}$, is finite. This can only be achieved with a finite Landau-Zener transition probability $p_\text{LZ}\neq0$.  Furthermore, we see that the inelastic relaxation that can occur between the two crossing events affects this occupation. Subsequently, the inelastic relaxation, from the left to the right level, decreases the left dot occupation, as it takes a finite time interval $\Delta\tilde{t}$ after the second crossing before the left dot can be discharged at $\epsilon_\text{L}>0$. Crucially, only the latter process is sensitive to the pumping direction. Taking the ratio with the reversed pumping current we obtain
\begin{equation}
\frac{\overline{I}_\text{C}}{\overline{I}_\text{C}'}=e^{-\Gamma_\text{in}(\Delta\tilde{t}-\Delta\tilde{t}')}\ .
\end{equation}
In triangle C  we can therefore directly readout the relaxation rate $\gamma_\text{in}$ from the excited (left) orbital state of the DQD to the ground (right) state. (Equivalently in the second triangle indicated in Fig.~\ref{fig_model}~(d) $\tilde{\gamma}_\text{in}$ can be extracted). This result also holds to a high precision beyond the approximation,  $\Gamma_\text{L},\Gamma_\text{R}\gg\Omega$, performed above for simplicity.

\section{Energy-dependent coupling strength}\label{sec_inelastic_E}

We have until now focussed on highly confined systems, such as single-dopant setups, where a readout of relaxation rates and tunnel coupling strengths has up to now posed problems. In these systems the inelastic relaxation rate from an excited state is energy-independent, $\Gamma_\text{in}=\text{const.}$ This is confirmed by experiments, where -- over the entire experimental range -- the dc leakage current between ground and first excited state was independent of the detuning and free of other parasitic effects, see Ref.~\cite{Roche12}. 
However,  in other cases -- as for example in DQDs realized in 2DEG systems with level spacings of the order of tenths of meV~\cite{Fujisawa98}  -- the internal, inelastic relaxation rate depends on the energy of the level spacing in a more complex manner than in the case discussed previously.  In Ref.~\cite{Fujisawa98} the inelastic rate has been extracted from a fit which in general is not independent of the tunneling rates to the electronic reservoirs, see App.~\ref{app_dc}.

Interestingly,  with the help of our method,  the rates $\gamma_\text{in}(E)$ and $\tilde{\gamma}_\text{in}(E)$ can still be extracted from plateaus in a fitting-free way, analogously to the cases shown in the previous Sections.~\footnote{Note that the readout of the charge relaxation rates due to tunneling to the reservoirs can possibly be inhibited in this case. The charge relaxation readout is however still possible if (1) the energy-dependent part of the inelastic relaxation rate is small, namely if $\delta\Gamma_\text{in}/\Gamma_\text{in}\ll 1$ where $\gamma_\text{in}(E)=(\Gamma_\text{in}+\delta\Gamma_\text{in}F(E))\theta(E)$ is the total inelastic rate, or if (2) the inelastic rate changes on an energy scale $\omega_\text{co}$ much larger than the driving amplitudes, $\delta\epsilon_\text{L,R}$.}   Also here, these plateau values depend on the inelastic rate, where we replace $\Gamma_\text{in}\rightarrow\Gamma_\text{in}F(E)$ with an energy-dependent function $F(E)$, and on externally fixed parameters, only.
The fitting procedure for the energy-dependent inelastic relaxation rate of Ref.~\cite{Fujisawa98}  requires $\Gamma_\text{in}$ to be the smallest coupling, $\Gamma_\text{in}\ll \Gamma_\text{L},\Gamma_\text{R}$. Here, we are not limited by this requirement, as we show in Sec.~\ref{sec_precision}. However, in the general case when the tunneling and the inelastic rate are comparable, the system undergoes a rather complex time evolution, because $\gamma_\text{in}$ and $\tilde{\gamma}_\text{in}$ are now \textit{time}-dependent through the time-dependent energy level positions. For the sake of simplicity, we hence first treat the limiting case of $\Omega,\Gamma_\text{in}\ll \Gamma_\text{L},\Gamma_\text{R}$. In this case, the current ratio of triangle C can be approximated as (see App.~\ref{app_triangleC2}),
\begin{equation}\label{eq_I_C_energy_dependent}
\frac{\overline{I}_\text{C}}{\overline{I}_\text{C}'}=e^{-\int_{\Delta\tilde{t}}dt\gamma_\text{in}(t)+\int_{\Delta\tilde{t}'}dt\gamma_\text{in}(t)}\ .
\end{equation}
An equivalent expression as a function of $\tilde{\gamma}_\text{in}$ is found from the lower triangle indicated in Fig.~\ref{fig_model}~(d). The current ratio contains information about the inelastic rate, time-averaged over the time intervals, $\Delta\tilde{t}$ and $\Delta\tilde{t}'$. Since the energy-level-spacing of the DQD is time-dependent, the time-averaging consequently corresponds to an average over a certain energy interval. Accordingly, it depends on the chosen trajectory to which extent an energy-resolved readout is possible. The time intervals $\Delta\tilde{t}$ and $\Delta\tilde{t}'$ can be made sufficiently small as compared to the time scale on which $\gamma_\text{in}$ effectively changes, by implementing convenient narrow, elliptic driving cycles in parameter space, e.g., by choosing the driving amplitudes as $\delta\epsilon_\text{R}\ll\delta\epsilon_\text{L}$. For this choice, Eq.~\eqref{eq_I_C_energy_dependent} can be approximated as
\begin{equation}\label{eq_inel_energy_dep}
\frac{\overline{I}_C}{\overline{I}_C'}\left(\overline{\epsilon}_\text{L},\overline{\epsilon}_\text{R}\right)\approx 1-\frac{2}{\Omega}\frac{\delta\epsilon_\text{R}}{\delta\epsilon_\text{L}}\gamma_\text{in}\left(-\overline{\epsilon}_\text{R}\right)\ .
\end{equation}
Here, we see explicitly that the current ratio is independent of the value of $\bar{\epsilon}_\text{L}$ for every fixed value of $\overline{\epsilon}_\text{R}$. If one  sweeps the average position of the right energy level, $\overline{\epsilon}_\text{R}$, one can thus achieve a reliable, energy-resolved readout of $\gamma_\text{in}(E)$ from a plateau (in $\epsilon_\text{L}$-direction) obtained at every energy value $\epsilon_\text{R}$. All other parameters occurring in Eq.~(\ref{eq_inel_energy_dep}) are input parameters of the driving scheme. In Fig.~\ref{fig_numeric}~(b), we demonstrate the applicability of this method at the example of $\gamma_\text{in}(\epsilon)=\theta(-\epsilon)\Gamma_\text{in}|\epsilon|\exp\left[-|\epsilon|\right]$ (and $\tilde{\gamma}_\text{in}(\epsilon)=\gamma_\text{in}(-\epsilon)$), with the dimensionless energy-variable $\epsilon=(\epsilon_\text{L}-\epsilon_\text{R})/\omega_\text{co}$. Such a rate corresponds to the coupling to an ohmic phonon bath in the limit of large amplitudes $\delta\epsilon_\text{L,R}\gg k_\text{B}T$, where the exponential cut-off with the cut-off frequency of the phonon bath $\omega_\text{co}$  is a standard procedure to describe the high energy regime, see, e.g., Ref.~\cite{Leggett87}.

In Fig.~\ref{fig_numeric}~(b), a sufficiently high $\Gamma\gg\Gamma_\text{in}$ has been chosen, fulfilling the requirements leading to Eq.~\eqref{eq_I_C_energy_dependent}, and has been compared to the case where $\Gamma$ and $\Gamma_\text{in}$ are of the same order of magnitude. The slight deviation of the extracted rate $\gamma_\text{in}^{\text{ex}}$ from the predicted one $\gamma_\text{in}$ occurring in the latter case is discussed in the following Sec.~\ref{sec_precision}.

\section{Precision of the readout scheme} \label{sec_precision}

\begin{figure}
\includegraphics[width=0.48\textwidth]{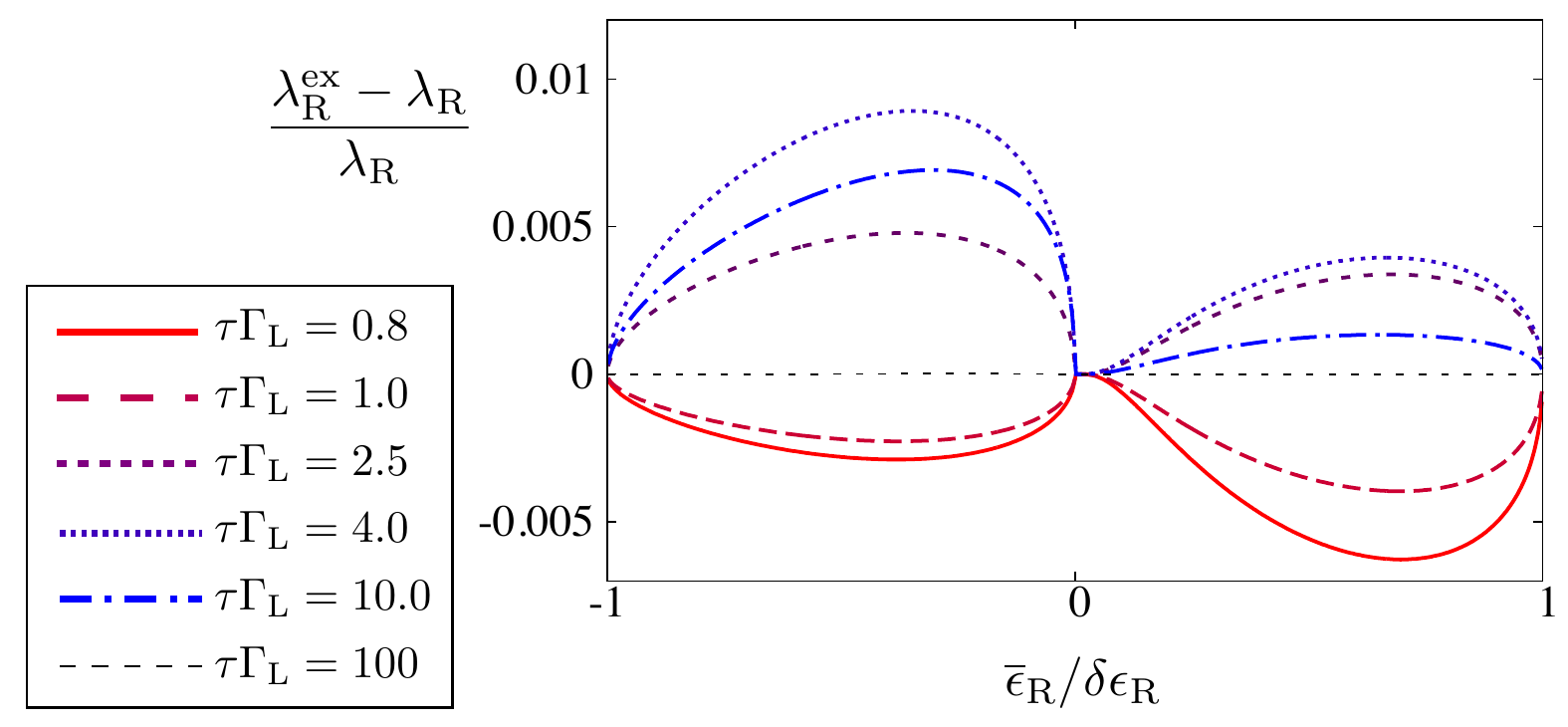}
\caption{Relative deviation of the extracted charge relaxation rate $\gamma_\text{R}^\text{ex}$ from the true rate $\gamma_\text{R}$, for different values of $\Gamma_\text{L}$ with $\Gamma_\text{R}=2.5\tau^{-1}$ and the remaining parameters as in Fig.~\ref{fig_trajectories}~(b).
}
\label{fig_precision}
\end{figure}

One of the strong points of our proposed readout scheme is the fact that different rates can be read out truly independently from plateau values within the different triangular regions, as indicated in Fig.~\ref{fig_model}~(d). 
The underlying equations for the charge current ratios providing the rate readout, are however only exact, if certain conditions concerning the relative magnitude of the different rates are fulfilled. In this Section we show that deviations that occur, when these conditions are not met, are negligibly small. More precisely, they can be shown to be mostly of the order of one percent. 

In Sec.~\ref{sec_charge}, we provide analytical formulas for the charge current and the charge current ratios necessary for the readout of the \textit{charge relaxation rate} $\gamma_\text{R}$, derived in the limit $\Gamma_\text{L}\gg\Gamma_\text{R}$. Deviations from the charge relaxation rate, $\gamma_\text{R}^\text{ex}$, extracted from the readout, from the true charge relaxation rate $\gamma_\text{R}$, were not visible for the parameters chosen for the plot in Fig.~\ref{fig_trajectories}~(b). In Fig.~\ref{fig_precision}, we show the relative deviation between  $\gamma_\text{R}^\text{ex}$ and $\gamma_\text{R}$, given by $(\gamma_\text{R}^\text{ex}-\gamma_\text{R})/\gamma_\text{R}$ for different values of $\Gamma_\text{L}$ and $\Gamma_\text{R}$ by zooming into the plateau region. One can see from the plot that as soon as $\Gamma_\text{L}\gg\Gamma_\text{R}$ is not fulfilled, the extracted rate is either slightly larger or slightly smaller than the actual rate, depending on the explicit values of $\Gamma_\text{L}$ and $\Gamma_\text{R}$. The behavior is not the same for triangle A ($\epsilon_\text{R}>0$) and for triangle B ($\epsilon_\text{R}>0$).  This clearly shows that the deviation from the ideal value is due to a complex interplay of tunneling, inelastic, and Landau-Zener transitions, which can occur as soon as $\Gamma_\text{L}\gg\Gamma_\text{R}$ is not given.
What is important, is that the maximal possible error is always mostly of the order of $1\%$. Similar results can be found for the precision of the readout of the energy-independent inelastic rate, $\Gamma_\text{in}$, in Sec.~\ref{sec_inelastic}, when the condition $\Gamma_\text{L,R}\gg\Omega$ is not fulfilled any longer.

\begin{figure}
\includegraphics[width=0.42\textwidth]{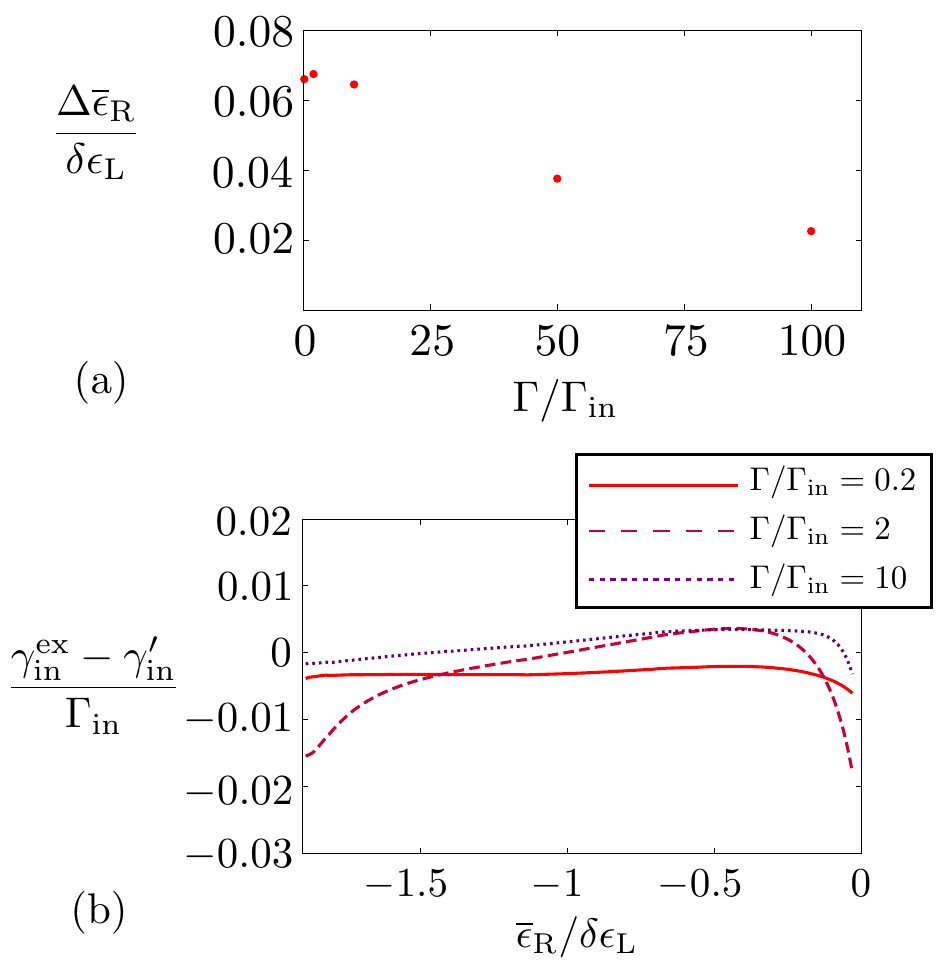}
\caption{(a) The energy shift $\Delta\overline{\epsilon}_\text{R}$ between the maxima of the  inelastic relaxation rate $\gamma_\text{in}^\text{ex}$ and the true rate $\gamma_\text{in}$ for several values of $\Gamma$. (b) Relative error of the extracted inelastic relaxation rate $\gamma_\text{in}^\text{ex}$ from the shifted rate $\gamma_\text{in}'$ for various values of $\Gamma$, such that the maxima of $\gamma_\text{in}^\text{ex}$ and $\gamma_\text{in}'$ occur at the same $\overline{\epsilon}_\text{R}$. The magnitude of the inelastic rate is $\Gamma_\text{in}=2\tau^{-1}$. The remaining parameters are as in Fig.~\ref{fig_numeric}~(b).
}
\label{fig_precision_in}
\end{figure}

In Sec.~\ref{sec_inelastic_E}, we illustrate how a possible \textit{energy dependence of the inelastic rate} can be extracted by means of Eq.~\eqref{eq_inel_energy_dep}. This equation is valid  in the regime where $\Gamma_\text{in}\ll\Gamma_\text{L,R}$ and $\delta\epsilon_\text{L}\gg\delta\epsilon_\text{R}$.  While the ratio of amplitudes is in general largely in the hands of the experimenter, the ratio between tunneling and inelastic rate may not be tunable. In the following, we demonstrate that for the readout of $\gamma_\text{in}(E),\tilde{\gamma}_\text{in}(E)$, we may well depart from $\Gamma_\text{in}\ll\Gamma_\text{L,R}$.

Figure~\ref{fig_numeric}~(b) shows the energy-dependence of the inelastic relaxation rate from the excited orbital DQD state and compares it to the extracted rate. Depending on the ratio $\Gamma/\tau^{-1}$, a slight shift occurs between them, while the shape of the energy-dependence is still well captured. We denote this shift, defined as the difference of the position of the maxima of $\gamma_\text{in}^\text{ex}$ and $\gamma_\text{in}$, by $\Delta\bar{\epsilon}_\text{R}$.  This shift depends on the ratio $\Gamma/\Gamma_\text{in}$, and does not exceed $0.07\delta\epsilon_\text{L}$ for the parameters chosen in Fig.~\ref{fig_precision_in}~(a). The result of the relative error when taking into account the shift $\Delta\overline{\epsilon}_\text{R}$ in $\gamma_\text{in}'(\epsilon)=\gamma_\text{in}(\epsilon+\Delta\overline{\epsilon}_\text{R})$ is shown in Fig.~\ref{fig_precision_in}~(b). We see that the error is of order $1\%$ even when $\Gamma\ll\Gamma_\text{in}$, the opposite case as compared to the one shown in Sec.~\ref{sec_inelastic_E}. Equation~\eqref{eq_inel_energy_dep} is hence applicable well beyond $\Gamma_\text{L,R}\gg\Gamma_\text{in}$. Therefore, a precise readout of a general inelastic relaxation rate is possible independently of the relative magnitude of tunneling and inelastic rates.

\section{Conclusion and Outlook} 

We provided a readout method to extract the charge relaxation rates as well as the relaxation rate of excited orbital states of a DQD by one single device-operation mode, based on nonadiabatic pumping.  By exploiting the sensitivity of the current to the direction of the pumping cycle, we showed that the ratio of currents of opposite pumping direction carries direct information on the characteristic relaxation rates of the system. 
Importantly, this information is contained in different plateaus of current ratios, in which the \textit{only free} parameter is the respective relaxation rate itself, yielding a fitting-free readout method.

We demonstrated the working principle of the readout method by deriving analytical formulas, which can be straightforwardly interpreted. In particular, they visualize the impact of different fast or slow relaxation processes on the magnitude of the pumping current. Subsequently, we fully release the conditions that were necessary for the analytical calculations and prove the reliability of the readout method numerically. This shows that, independently of the relative size of coupling constants a readout  is possible with an error that is maximally of the order of $1\%$. 

The exploitation of this method in experiment is expected to be possible in different realizations of gated DQDs, see e.g. Refs.~\cite{Roche13}. Beyond this, our method can be generalized for other types of devices, in which time-dependent local control over discrete energy levels allows for the implementation of the proposed relaxation-rate readout scheme.

\acknowledgments
We are grateful for helpful discussions with M. Sanquer, C. Stampfer, and M. Misiorny. This work has been financially supported by the Knut and Alice Wallenberg foundation through the Wallenberg Academy Fellows program (J.~S.), by the Ministry of Innovation NRW, Germany (J.~S. and R.-P.~R.), by the EU through the FP7 ICT collaborative project SiAM, Project No. 610637 (X.~J.), and by the Nanosciences Foundation in Grenoble, in the frame of its Chair of Excellence program (R.-P.~R. and X.~J.).

\appendix

\section{Rate-readout from dc measurement}\label{app_dc}

\subsection{Voltage biased current and rate readout scheme}
In the case where transport occurs through two single levels at energies $\epsilon_\text{L}$ and $\epsilon_\text{R}$, and no inelastic internal transitions in the DQD occur, the stationary current due to a voltage bias $V$ is given by~\cite{Nazarov93,Stoof96}:
\begin{equation}
 \frac{I_\text{dc}}{e}= \frac{\Delta^2 \Gamma_\text{L}}{\Delta^2 (2 + \frac{\Gamma_\text{L}}{\Gamma_\text{R}})+ \frac{\Gamma_\text{L}^2}{4} + \frac{(\epsilon_\text{L}-\epsilon_\text{R})^2}{\hbar^2}}.
\label{eq_dc}
\end{equation}
Note that $I_\text{dc}$ does not depend on $V$, as long as $V$ remains larger than the intrinsic energy width of the levels and temperature. This expression is valid for the lower triple point. For the upper triple point, $\Gamma_\text{L}$ and $\Gamma_\text{R}$ must be swapped. Equation~(\ref{eq_dc}) however contains many unknowns and hence some assumptions on the relative magnitude of the rates have to be made to extract them. 
If also inelastic internal hopping processes occur, an additional contribution to the dc current arises~\cite{Fujisawa98}, which - in contrast to Eq.~\eqref{eq_dc} - is asymmetric in $\epsilon_\text{L}-\epsilon_\text{R}$ due to an asymmetric energy dependence of $\Gamma_\text{in}F(E)$. Note that this asymmetry stems from the imposed voltage bias direction and does not apply to the equilibrium situation discussed in the main paper. The asymmetry yields a possibility to distinguish the two current contributions. The inelastic current contribution is given by
\begin{equation}
I_\text{dc}^\text{in}=\frac{e}{\Gamma_\text{L}^{-1}+\Gamma_\text{R}^{-1}+(\Gamma_\text{in}F(E))^{-1}}.
\label{eq_dc_in}
\end{equation}
For $\Gamma_\text{L,R}\gg\Gamma_\text{in}$, this expression reduces to $I_\text{dc}^\text{in}=e\Gamma_\text{in}F(E)$, allowing for a readout of $\Gamma_\text{in}F(E)$.
The combination of Eqs.~(\ref{eq_dc}) and (\ref{eq_dc_in}) allows for the extraction of $\Gamma_\text{L}$, $\Gamma_\text{L}$, $\Delta$, and $\Gamma_\text{in}F(E)$ from a complex fitting procedure. 

\subsection{Comparison of readout speed}

Our proposed readout scheme involves the measurement of a time-averaged current due to time-dependent driving. Here we show that the measurement time is comparable to the one of a readout with stationary voltages.

In general, the lower bound of the time needed to average the current signal is given by the signal-to-noise ratio,
$T_0>S/\overline{I}^2$,
where the current noise at the detector is defined as $S$, and $I=I_\text{dc}^\text{in}$ for the voltage-biased case and $I=\overline{I}$ for the pumping case. The leading noise contribution is dominated by external sources, and highly sensitive on the details of the experimental setup. On the other hand, it depends very little on the specific operation, which is why we assume that $S$ is roughly of the same order of magnitude for the voltage-biased case as for the time-dependent electron pump, when investigating the same system. 

In the voltage-biased case, the current is dominated by the smallest rate, in Ref.~\cite{Fujisawa98} the inelastic rate, i.e., $I\sim e\Gamma_\text{in}$. For the electron pump, for the triangular regions of interest, $I\sim e\Omega p_\text{err}$, with an error probability $p_\text{err}$ due to nonadiabatic driving. In the strongly nonadiabatic driving regime, where $\Omega$ is of the same order or larger than the relaxation rates, this error is of order 1, hence $I\sim e\Omega$. For weakly nonadiabatic driving on the other hand, the current in the triangles is exponentially suppressed. 
Therefore, as soon as the  driving is sufficiently fast such as prescribed by our method, the mere averaging time, required for our scheme does not exceed the one of the dc case.

\section{Analytic current expression for the charge relaxation rate readout at triangle A} \label{app_triangleA}
In this Section, we derive the expression  Eq.~\eqref{eq_I_A} for the current in triangle
A (see, e.g., Fig.~\ref{fig_model}~(c)), in the limit of a large tunnel coupling to the left dot, $\Gamma_{\text{L}}\gg\Gamma_{\text{R}},\Gamma_{\text{in}}$. As shown in Fig.~\ref{fig_trajectories}~(a), the time evolution of the energy levels $\epsilon_{\text{L,R}}$
exhibits four different configurations in the pumping parameter
space. Within each of these intervals, the kernel is a constant
function of time, since we can neglect the energy-dependence of the Fermi functions due to the large-amplitude driving. We are therefore able to solve the time evolution of the density matrix piecewise, with the corresponding kernels of
these four configurations. For the charging of the right dot $\epsilon_{\text{L}}>0$,
$\epsilon_{\text{R}}<0$ during time interval $\Delta\tau_{\text{c}}$, we have
\begin{eqnarray}
W\left(t\in \Delta\tau_\text{c}\right)=:W_{\text{c}} =\left(\begin{array}{ccc}
-2\Gamma_{\text{R}} & \Gamma_{\text{L}} & 0\\
0 & -\Gamma_{\text{L}}-\Gamma_{\text{in}} & 0\\
2\Gamma_{\text{R}} & \Gamma_{\text{in}} & 0
\end{array}\right) .
\end{eqnarray}
For the subsequent discharging when $\epsilon_{\text{R}}>0$,
we have three ``sub-stages'': first when $\epsilon_{\text{L}}>\epsilon_{\text{R}}$
during $\Delta t$, 
\begin{eqnarray}
W\left(t\in\Delta t\right)=:W_{\text{d}} =\left(\begin{array}{ccc}
0 & \Gamma_{\text{L}} & \Gamma_{\text{R}}\\
0 & -\Gamma_{\text{L}}-\Gamma_{\text{in}} & 0\\
0 & \Gamma_{\text{in}} & -\Gamma_{\text{R}}
\end{array}\right), 
\end{eqnarray}
and secondly after the first level crossing, where
$\epsilon_{\text{L}}<\epsilon_{\text{R}}$ during $\Delta\tau$,
\begin{eqnarray}\label{eq_KernelDeltat}
W\left(t\in\Delta\tau\right)=:W_{\overline{\text{d}}} & =\left(\begin{array}{ccc}
0 & \Gamma_{\text{L}} & \Gamma_{\text{R}}\\
0 & -\Gamma_{\text{L}} & \Gamma_{\text{in}}\\
0 & 0 & -\Gamma_{\text{R}}-\Gamma_{\text{in}}
\end{array}\right).
\end{eqnarray}
 Finally, after the second level crossing we again have $\epsilon_{\text{L}}>\epsilon_{\text{R}}$
during $\Delta t'$, leading to the same Kernel as given in Eq.~(\ref{eq_KernelDeltat}), $W\left(t\in\Delta t'\right)=W\left(t\in\Delta t\right)=W_\text{d}$.

In order to compute the time evolution of the density matrix, we provide
the propagators for each configuration, defined as $\Pi_{x}=e^{W_{x}t}$,
with $x\in\left\{ \text{c},\text{d},\overline{\text{d}}\right\} $,
\begin{eqnarray}
\label{eq_prop_A_1}\Pi_{\text{c}}\left(t\right) & = & \left(\begin{array}{ccc}
0 & 0 & 0\\
0 & 0 & 0\\
1 & 1 & 1
\end{array}\right)+e^{-2\Gamma_{\text{R}}t}\left(\begin{array}{ccc}
1 & \frac{\Gamma_\text{L}}{\Gamma_\text{in}+\Gamma_\text{L}-2\Gamma_\text{R}} & 0\\
0 & 0 & 0\\
-1 & -\frac{\Gamma_\text{L}}{\Gamma_\text{in}+\Gamma_\text{L}-2\Gamma_\text{R}} & 0
\end{array}\right)\nonumber\\
&&+e^{-(\Gamma_\text{in}+\Gamma_\text{L})t}
\left(\begin{array}{ccc}
0 & - \frac{\Gamma_\text{L}}{\Gamma_\text{in}+\Gamma_\text{L}-2\Gamma_\text{R}} &0\\
0&1&0\\
0& -\frac{\Gamma_\text{in}-2\Gamma_\text{R}}{\Gamma_\text{in}+\Gamma_\text{L}-2\Gamma_\text{R}} &0
\end{array}\right)
\\ \label{eq_prop_A_2}
\Pi_{\text{d}}\left(t\right) & = & \left(\begin{array}{ccc}
1 & 1 & 1\\
0 & 0 & 0\\
0 & 0 & 0
\end{array}\right)
+e^{-\Gamma_{\text{R}}t}\left(\begin{array}{ccc}
0 & -\frac{\Gamma_\text{in}}{\Gamma_\text{in}+\Gamma_\text{L}-\Gamma_\text{R}} & -1\\
0 & 0 & 0\\
0 &  \frac{\Gamma_\text{in}}{\Gamma_\text{in}+\Gamma_\text{L}-\Gamma_\text{R}} & 1
\end{array}\right)\nonumber\\
&&+ e^{-(\Gamma_\text{in}+\Gamma_\text{L})}
\left(\begin{array}{ccc}
0&-\frac{\Gamma_\text{L}-\Gamma_\text{R}}{\Gamma_\text{in}+\Gamma_\text{L}-\Gamma_\text{R}}&0\\
0&1&0\\
0& -\frac{\Gamma_\text{in}}{\Gamma_\text{in}+\Gamma_\text{L}-\Gamma_\text{R}}&0
\end{array}\right)
\\ \label{eq_prop_A_3}
\Pi_{\overline{\text{d}}}\left(t\right) & = & \left(\begin{array}{ccc}
1 & 1 & 1\\
0 & 0 & 0\\
0 & 0 & 0
\end{array}\right)+e^{-\left(\Gamma_{\text{R}}+\Gamma_{\text{in}}\right)t}\left(\begin{array}{ccc}
0 & 0 & \frac{\Gamma_\text{L}-\Gamma_\text{R}}{\Gamma_\text{in}-\Gamma_\text{L}+\Gamma_\text{R}}\\
0 & 0 & -\frac{\Gamma_\text{in}}{\Gamma_\text{in}-\Gamma_\text{L}+\Gamma_\text{R}}\\
0 & 0 & 1
\end{array}\right)\nonumber\\
&&+e^{-\Gamma_\text{L}t}
\left(\begin{array}{ccc}
0 & -1 & -\frac{\Gamma_\text{in}}{\Gamma_\text{in}-\Gamma_\text{L}+\Gamma_\text{R}}\\
0 & 1 & \frac{\Gamma_\text{in}}{\Gamma_\text{in}-\Gamma_\text{L}+\Gamma_\text{R}}\\
0&0&0
\end{array}\right)
.
\end{eqnarray}
In addition we have to provide the propagator for the Landau-Zener
level crossing,
\begin{equation}
\Pi_{\text{LZ}}=\left(\begin{array}{ccc}
1 & 0 & 0\\
0 & p_{\text{LZ}} & 1-p_{\text{LZ}}\\
0 & 1-p_{\text{LZ}} & p_{\text{LZ}}
\end{array}\right).
\end{equation}
With these expressions, we can construct the propagator over one pumping period $\tau$. Setting the initial time $t_{0}$ as the time at which $\epsilon_{\text{R}}$ changes from negative to positive (that is, at the end of the charging
process taking place in the interval $\Delta\tau_\text{c}$), the full propagator is given as
\begin{eqnarray}
&& \Pi\left(\tau+t_{0},t_{0}\right)=\\
& & \Pi_{\text{c}}\left(\Delta\tau_{\text{c}}\right)\Pi_{\text{d}}\left(\Delta t'\right)\Pi_{\text{LZ}}\Pi_{\overline{\text{d}}}\left(\Delta\tau\right)\Pi_{\text{LZ}}\Pi_{\text{d}}\left(\Delta t\right).\nonumber
\end{eqnarray}
In general, the time evolution may start with some arbitrary initial
density matrix. For any
initial density matrix, the steady state solution of the
density matrix, $P_{\text{ss}}$, is reached  after a sufficient number of pumping cycles
(roughly speaking, the number of cycles must be much larger than $\frac{1}{\Gamma_{\text{R}}\tau},\frac{1}{\Gamma_{\text{in}}\tau}$).
A steady state density matrix is found if it satisfies the boundary condition
\begin{equation}\label{eq_steady}
P_{\text{ss}}\left(t\right)=\Pi\left(\tau+t,t\right)P_{\text{ss}}\left(t\right),
\end{equation}
for any time $t$.
When plugging in the propagators above, the steady state density matrix at $t_0$
right after the charging is found to be
\begin{equation}
P_{\text{ss}}\left(t_{0}\right)=\left(\begin{array}{c}
1-n^{\text{R}}\\
0\\
n^{\text{R}}
\end{array}\right). 
\end{equation}
Here, we took the limit of $\Gamma_{\text{L}}\gg\Gamma_{\text{R}},\Gamma_{\text{in}}$, approximating the fast time evolution with rate $\Gamma_\text{L}$ to be quasi instantaneous, $e^{-\Gamma_\text{L}t}\rightarrow 0$, and neglecting terms of the order $1/\Gamma_\text{L}$. Furthermore, 
\begin{equation}
n^{\text{R}}=\frac{1-e^{-2\Gamma_{\text{R}}\Delta\tau_{\text{c}}}}{1-e^{-\Gamma_{\text{R}}\Delta t'}e^{-\left(\Gamma_{\text{R}}+\Gamma_{\text{in}}\right)\Delta\tau}e^{-2\Gamma_{\text{R}}\Delta\tau_{\text{c}}}e^{-\Gamma_{\text{R}}\Delta t}p_{\text{LZ}}^{2}},
\end{equation}
is the steady state occupation of the right dot immediately after 
the charging. It can be transformed into the formula for $n^{\text{R}}$ given in the main text right after Eq.~\eqref{eq_I_A}
by using the identity, $\tau=\Delta t+\Delta\tau+\Delta t'+\Delta\tau_{\text{c}}$.
After this transformation one immediately sees that $n^{\text{R}}$ does not depend on $\Delta t,\Delta t'$.
The time evolution of the density matrix is fully determined with
the steady state density matrix and the propagators for each configuration.

This allows us to proceed to the calculation of the current. For the
current, we need the object, $W_{I}$, which is derived from the kernel
$W$ by retaining all elements
which change the total charge of the DQD, and adding
the sign according to whether the charge leaves/enters through the left or right
contact. Likewise, this object is piecewise constant in time, where
the different parts are given as
\begin{eqnarray}
W_{I,\text{c}}  =  \frac{1}{2}\left(\begin{array}{ccc}
0 & -\Gamma_{\text{L}} & 0\\
0 & 0 & 0\\
-2\Gamma_{\text{R}} & 0 & 0
\end{array}\right)\\
W_{I,\text{d}}  = W_{I,\overline{\text{d}}}  =  \frac{1}{2}\left(\begin{array}{ccc}
0 & -\Gamma_{\text{L}} & \Gamma_{\text{R}}\\
0 & 0 & 0\\
0 & 0 & 0
\end{array}\right).
\end{eqnarray}
The factor $1/2$ in front comes from the fact that we choose to consider
the symmetrised current $I=\left(I_\text{L}-I_\text{R}\right)/2$. (This choice
is made just for convenience, as for the dc component of the current,
it actually makes no difference whether the current is symmetrised
or not.) With above time evolution of the density matrix, and the
object $W_{I}$, we are able to calculate the dc current $I=\int_{0}^{\tau}\frac{dt}{\tau}I\left(t\right)$,
as follows,
\begin{eqnarray}\label{eq_current_piecewise_A}
\frac{\tau I}{e} & = & \int_{0}^{\Delta t}dte^{T}W_{I,\text{d}}\Pi_{\text{d}}\left(t\right)P_{\text{ss}}\left(t_{0}\right)\label{eq_current_propagators}\\
& &+\int_{0}^{\Delta\tau}dte^{T}W_{I,\overline{\text{d}}}\Pi_{\overline{\text{d}}}\left(t\right)P_{\text{ss}}\left(t_{0}+\Delta t\right)\nonumber\\
 &  & +\int_{0}^{\Delta t'}dte^{T}W_{I,\text{d}}\Pi_{\text{d}}\left(t\right)P_{\text{ss}}\left(t_{0}+\Delta t+\Delta\tau\right)\nonumber\\
 &  & +\int_{0}^{\Delta\tau_{\text{c}}}dte^{T}W_{I,\text{c}}\Pi_{\text{c}}\left(t\right)P_{\text{ss}}\left(t_{0}+\Delta t+\Delta\tau+\Delta t'\right).\nonumber
\end{eqnarray}
Here, $e^{T}=\left(1,1,1\right)$ represents the trace operator and
\begin{eqnarray*}
P_{\text{ss}}\left(t_{0}+\Delta t\right) & = & \Pi_{\text{LZ}}\Pi_{\text{d}}\left(\Delta t\right)P_{\text{ss}}\left(t_{0}\right)\\
P_{\text{ss}}\left(t_{0}+\Delta t+\Delta\tau\right) & = & \Pi_{\text{LZ}}\Pi_{\overline{\text{d}}}\left(\Delta\tau\right)P_{\text{ss}}\left(t_{0}+\Delta t\right)\\
P_{\text{ss}}\left(t_{0}+\Delta t+\Delta\tau+\Delta t'\right) & = & \Pi_{\text{d}}\left(\Delta t'\right)P_{\text{ss}}\left(t_{0}+\Delta t+\Delta\tau\right).
\end{eqnarray*}
Note that it is important that the propagators in the current expression, Eq.~(\ref{eq_current_propagators}), are the full expressions of Eqs.~(\ref{eq_prop_A_1}) to (\ref{eq_prop_A_3}) and that the limit $\Gamma_\text{L}\gg \Gamma_\text{R},\Gamma_\text{in}$ can only be taken after the time integral has been evaluated. The reason for this is that small terms scaling with $\Gamma_\text{L}^{-1}$ or $e^{-\Gamma_\text{L}t}$ in the propagators are compensated by terms in the current kernels $W_{I,x}$ that scale with $\Gamma_\text{L}$. 

After a little algebra, Eq.~(\ref{eq_current_piecewise_A}) leads to the result given in the main text, Eq.~\eqref{eq_I_A}.

\section{Analytic current expression for the charge relaxation rate readout at triangle C}\label{app_triangleC}

\subsection{Energy-independent inelastic relaxation rate}\label{app_triangleC1}
The calculation for the current in triangle C, Eq.~\eqref{eq_current_triangle_C_const}, with a constant inelastic relaxation rate, can be made
in similar terms as the one for A. We here have the piecewise constant kernels 
\begin{eqnarray}
&&W\left(t\in\Delta\tilde{t},\Delta\tilde{t'}\right)=:W_{\text{c}}  =\left(\begin{array}{ccc}
-2\Gamma_{\text{L}}-2\Gamma_{\text{R}} & 0 & 0\\
2\Gamma_{\text{L}} & -\Gamma_{\text{in}} & 0\\
2\Gamma_{\text{R}} & \Gamma_{\text{in}} & 0
\end{array}\right)\nonumber\\ \label{eq:w_c}
\end{eqnarray}
\begin{eqnarray}
&&W\left(t\in\Delta\tilde{\tau}_\text{c}\right)=:W_{\text{d}}=\left(\begin{array}{ccc}
-2\Gamma_{\text{R}} & \Gamma_{\text{L}} & 0\\
0 & -\Gamma_{\text{L}}-\Gamma_{\text{in}} & 0\\
2\Gamma_{\text{R}} & \Gamma_{\text{in}} & 0
\end{array}\right) \label{eq:w_d}
\end{eqnarray}
\begin{eqnarray}
&&W\left(t\in\Delta\tilde{\tau}\right)=:W_{\overline{\text{c}}} =\left(\begin{array}{ccc}
-2\Gamma_{\text{L}}-2\Gamma_{\text{R}} & 0 & 0\\
2\Gamma_{\text{L}} & 0 & \Gamma_{\text{in}}\\
2\Gamma_{\text{R}} & 0 & -\Gamma_{\text{in}}
\end{array}\right) ,\label{eq:w_cbar}
\end{eqnarray}
with the corresponding propagators $\Pi_{\text{c}}\left(t\right), \Pi_{\text{d}}\left(t\right)$ and $\Pi_{\overline{\text{c}}}\left(t\right)$. The full propagator for one pumping period is
\begin{eqnarray}\label{eq_full_prop_C}
&&\Pi\left(\tau+\widetilde{t}_{0},\widetilde{t}_{0}\right)=\\
&&\Pi_{\text{LZ}}\Pi_{\overline{\text{c}}}\left(\Delta\widetilde{\tau}\right)\Pi_{\text{LZ}}\Pi_{\text{c}}\left(\Delta\widetilde{t}'\right)\Pi_{\text{d}}\left(\Delta\tilde{\tau}_{\text{c}}\right)\Pi_{\text{c}}\left(\Delta\tilde{t}\right)\nonumber,
\end{eqnarray}
where we indicate the time right after the second crossing as $\widetilde{t}_{0}$.
Now, in the limit $\Gamma_{\text{L}},\Gamma_\text{R}\gg\Gamma_{\text{in}}$ and $e^{-\Gamma_\text{L,R}t}\rightarrow0$, the steady state condition, Eq.~(\ref{eq_steady}), together with Eq.~(\ref{eq_full_prop_C}) delivers
\begin{equation}
P_{\text{ss}}\left(\widetilde{t}_{0}\right)=\left(\begin{array}{c}
0\\
n_{\text{cr}}^{\text{L}}\\
1-n_{\text{cr}}^{\text{L}}
\end{array}\right)
\end{equation}
where $n_{\text{cr}}^{\text{L}}=p_{\text{LZ}}\left[1+e^{-\Gamma_{\text{in}}\Delta\widetilde{\tau}}\left(1-2p_{\text{LZ}}\right)\right]$,
is the occupation probability of the left dot right after the second
crossing, at the end of the interval $\Delta\tilde{\tau}$. In the case here, the current calculation is simpler. As
one can easily convince oneself, there is no current contribution
when both levels are below the Fermi level: there is no finite probability of an empty DQD for the time interval where
$\epsilon_{\text{L,R}}<0$; therefore, no charge transfer
between DQD system and reservoirs occurs. The only contribution to the current then
comes from the discharging of the left dot during the time interval
$\Delta\widetilde{\tau}_{\text{c}}$, which is given as
\begin{equation}\label{eq_current_C}
\frac{\tau I}{e}=\int_{0}^{\Delta\widetilde{\tau}_{\text{c}}}dt'e^{T}W_{I\text{d}}\Pi_{\text{d}}\left(t'\right)P_{\text{ss}}\left(\Delta\tilde{t}+\widetilde{t}_{0}\right)
\end{equation}
with
\begin{equation}\label{eq_kernel_C}
W_{I\text{d}}=\frac{1}{2}\left(\begin{array}{ccc}
0 & -\Gamma_{\text{L}} & 0\\
0 & 0 & 0\\
-2\Gamma_{\text{R}} & 0 & 0
\end{array}\right)
\end{equation}
and
\begin{equation}\label{eq_SS_C}
P_{\text{ss}}\left(\Delta\tilde{t}+\widetilde{t}_{0}\right)=\Pi_{\text{c}}\left(\Delta\tilde{t}\right)P_{\text{ss}}\left(\widetilde{t}_{0}\right)\approx\left(\begin{array}{c}
0\\
e^{-\Gamma_{\text{in}}\Delta\widetilde{t}}n_{\text{cr}}^{\text{L}}\\
1-e^{-\Gamma_{\text{in}}\Delta\widetilde{t}}n_{\text{cr}}^{\text{L}}
\end{array}\right).
\end{equation}
Inserting Eq.~(\ref{eq_kernel_C}) and (\ref{eq_SS_C}) into Eq.~(\ref{eq_current_C}), and again making the approximations $\Gamma_{\text{L,R}}\gg\Gamma_{\text{in}}$ and $e^{-\Gamma_\text{L,R}t}\rightarrow0$ only after performing the integral, we find Eq.~(\ref{eq_current_triangle_C_const}).

\subsection{Energy-dependent inelastic relaxation rate}\label{app_triangleC2}
We now derive the current expression for an energy-dependent $\gamma_{\text{in}},\widetilde{\gamma}_{\text{in}}$, see Eq.~\eqref{eq_I_C_energy_dependent}. Let us first comment on the general shape
of the energy dependence of the inelastic processes. The coupling to a bosonic bath (e.g. phonons) induces internal transitions, namely hopping between the two dot levels. Therefore, the inelastic rates can be written
as functions of the energy detuning $\epsilon_{\text{L}}-\epsilon_{\text{R}}$.
Furthermore, assuming that the memory of the reservoir decays very
fast (quasi instantanteously), the time dependence of the detuning
enters simply as a parameter, $\gamma_{\text{in}}\left(\epsilon_{\text{L}}\left(t\right)-\epsilon_{\text{R}}\left(t\right)\right),\widetilde{\gamma}_{\text{in}}\left(\epsilon_{\text{L}}\left(t\right)-\epsilon_{\text{R}}\left(t\right)\right)$.
Since in general, the kernel is then explicitly time dependent even
within the time intervals where the charge does not change, the time evolution of the density matrix becomes highly
complex. However, we may proceed with the following further simplifications. For a reservoir
in equilibrium, the forward and backward processes fulfill detailed
balance, $\gamma_{\text{in}}\left(\epsilon_{\text{L}}-\epsilon_{\text{R}}\right)/\widetilde{\gamma}_{\text{in}}\left(\epsilon_{\text{L}}-\epsilon_{\text{R}}\right)=e^{\left(\epsilon_{\text{L}}-\epsilon_{\text{R}}\right)/k_{\text{B}}T}$.
We now make use of the fact that the amplitudes are much larger
than the temperature broadening, in the regime considered here. We therefore find that depending on whether
$\epsilon_{\text{L}}\lessgtr\epsilon_{\text{R}}$ either the rate
$\gamma_{\text{in}}$ or $\widetilde{\gamma}_{\text{in}}$ is exponentially
suppressed (equivalent to a very cold reservoir).

Moreover, we again consider the limit $\Gamma_{\text{L,R}}\gg\gamma_{\text{in}},\widetilde{\gamma}_{\text{in}}$ for simplicity
(in analogy to the above case of energy-independent relaxation), where the
propagators can be approximated as follows. First we deal with the simplest
case, namely, the time evolution in the configuration of $\epsilon_{\text{L}}>0,\epsilon_{\text{R}}<0$.
This time evolution is described by kernel $W_{\text{d}}$, whose eigenvalues are given for each time $t$ as $-2\Gamma_{\text{R}}$
and $-\Gamma_{\text{L}}-\gamma_{\text{in}}\left(t\right)$. Hence
there is no slow relaxation rate, and the
corresponding propagator may be approximated in the limit $\Gamma_\text{L,R}\gg\gamma_\text{in}$,
\begin{equation}\label{eq_Prop_d_time_independent}
\begin{split}
\Pi_\text{d}(t)=\left(\begin{array}{ccc}
0 & 0 & 0\\
0 & 0 & 0\\
1 & 1 & 1
\end{array}\right)+e^{-2\Gamma_{\text{R}}t}\left(\begin{array}{ccc}
1 & \frac{\Gamma_{\text{L}}}{\Gamma_{\text{L}}-2\Gamma_{\text{R}}} & 0\\
0 & 0 & 0\\
-1 & -\frac{\Gamma_{\text{L}}}{\Gamma_{\text{L}}-2\Gamma_{\text{R}}} & 0
\end{array}\right)\\+e^{-\Gamma_{\text{L}}t}\left(\begin{array}{ccc}
0 & -\frac{\Gamma_{\text{L}}}{\Gamma_{\text{L}}-2\Gamma_{\text{R}}} & 0\\
0 & 1 & 0\\
0 & \frac{2\Gamma_{\text{R}}}{\Gamma_{\text{L}}-2\Gamma_{\text{R}}} & 0
\end{array}\right)\ .
\end{split}
\end{equation}
For the remaining two configurations, this is no longer true as their corresponding kernels contain one slow relaxation rate. Here we have to proceed as follows.
We rewrite the differential equations for these two configurations as
\begin{eqnarray}
\dot{P}(t\in\Delta\tilde{t},\Delta\tilde{t}')= & \left[W_{\text{c}}^{\text{T}}+\left(\begin{array}{ccc}
0 & 0 & 0\\
0 & -\gamma_{\text{in}}\left(t\right) & 0\\
0 & \gamma_{\text{in}}\left(t\right) & 0
\end{array}\right)\right]P\nonumber\\
\\
\dot{P}(t\in\Delta\tilde{\tau})= & \left[W_{\overline{\text{c}}}^{\text{T}}+\left(\begin{array}{ccc}
0 & 0 & 0\\
0 & 0 & \widetilde{\gamma}_{\text{in}}\left(t\right)\\
0 & 0 & -\widetilde{\gamma}_{\text{in}}\left(t\right)
\end{array}\right)\right]P.\nonumber\\
\end{eqnarray}
Here, $W_{x}^{\text{T}}=W_{x}\left(\Gamma_{\text{in}}=0\right)$ is
simply the tunneling part of the piecewise constant kernels as previously
defined in Eqs.~(\ref{eq:w_c}) and (\ref{eq:w_cbar}).
If the tunneling dynamics is much faster than the internal inelastic relaxation, one can envisage a separation
of time scales in the following manner. First, we make the transform
$P'=e^{-W_{x}^{\text{T}}t}P$ through which, in the limit $\Gamma_{\text{L,R}}\gg\gamma_{\text{in}},\widetilde{\gamma}_{\text{in}}$,
we arrive at the equations,
\begin{eqnarray}
\dot{P}'(t\in\Delta\tilde{t},\Delta\tilde{t}') & = & -\gamma_{\text{in}}\left(t\right)\left(\begin{array}{ccc}
0 & 0 & 0\\
\frac{\Gamma_{\text{L}}}{\Gamma_{\text{L}}+\Gamma_{\text{R}}} & 1 & 0\\
-\frac{\Gamma_{\text{L}}}{\Gamma_{\text{L}}+\Gamma_{\text{R}}} & -1 & 0
\end{array}\right)P'\nonumber\\
\\
\dot{P}'(t\in\Delta\tilde{\tau})& = & -\widetilde{\gamma}_{\text{in}}\left(t\right)\left(\begin{array}{ccc}
0 & 0 & 0\\
-\frac{\Gamma_{\text{R}}}{\Gamma_{\text{L}}+\Gamma_{\text{R}}} & 0 & -1\\
\frac{\Gamma_{\text{R}}}{\Gamma_{\text{L}}+\Gamma_{\text{R}}} & 0 & 1
\end{array}\right)P'.\nonumber\\
\end{eqnarray}
In this basis, we find thus an exactly solvable dynamics, because
the explicit time dependence appears only in the prefactor. Due to this explicit time-dependence it is now not sufficient anymore to define propagators of time-differences, but the propogators depend explicitly on two times. Eventually,
 we find the following solutions for the propagators of the untransformed $P$, in the limit $\Gamma_\text{L,R}\gg\gamma_\text{in}$ and $e^{-\Gamma_\text{L,R}t}\rightarrow 0$,
\begin{eqnarray}
\Pi_{\text{c}}\left(t,t'\right) & = & \left(\begin{array}{ccc}
0 & 0 & 0\\
0 & 0 & 0\\
1 & 1 & 1
\end{array}\right)\\
&&+e^{-\int_{t'}^{t}dt''\gamma_{\text{in}}\left(t''\right)}\left(\begin{array}{ccc}
0 & 0 & 0\nonumber\\
\frac{\Gamma_{\text{L}}}{\Gamma_{\text{L}}+\Gamma_{\text{R}}} & 1 & 0\\
-\frac{\Gamma_{\text{L}}}{\Gamma_{\text{L}}+\Gamma_{\text{R}}} & -1 & 0
\end{array}\right)\\
\Pi_{\overline{\text{c}}}\left(t,t'\right) & = & \left(\begin{array}{ccc}
0 & 0 & 0\\
1 & 1 & 1\\
0 & 0 & 0
\end{array}\right)\\
&&+e^{-\int_{t'}^{t}dt''\widetilde{\gamma}_{\text{in}}\left(t''\right)}\left(\begin{array}{ccc}
0 & 0 & 0\\
-\frac{\Gamma_{\text{R}}}{\Gamma_{\text{L}}+\Gamma_{\text{R}}} & 0 & -1\\
\frac{\Gamma_{\text{R}}}{\Gamma_{\text{L}}+\Gamma_{\text{R}}} & 0 & 1
\end{array}\right).\nonumber
\end{eqnarray}
In order to find the steady state, we insert the propagators into Eq.~\eqref{eq_full_prop_C}, and we obtain
\begin{equation}
P_{\text{ss}}\left(\widetilde{t}_{0}\right)=\left(\begin{array}{c}
0\\
n_{\text{cr}}^{\text{L}}\\
1-n_{\text{cr}}^{\text{L}}
\end{array}\right)
\end{equation}
with $n_{\text{cr}}^{\text{L}}=p_{\text{LZ}}\left[1+e^{-\int_{\Delta\widetilde{\tau}}dt'\widetilde{\gamma}_{\text{in}}\left(t'\right)}\left(1-2p_{\text{LZ}}\right)\right]$, where the notation $\int_{\Delta\widetilde{\tau}}dt'$
stands for an integral over the interval $t'\in\Delta\widetilde{\tau}$.
We are left with calculating the current expression, which, as in the case of a time-independent $\gamma_{\text{in}},\widetilde{\gamma}_{\text{in}}$, contains only a nonzero current contribution within the time evolution with $W_\text{d}$.
Then, using the propagator from Eq.~\eqref{eq_Prop_d_time_independent},  the final result for the current is computed according to Eq.~\eqref{eq_current_C}, and amounts to
\begin{equation}
\frac{\tau I}{e}=-e^{-\int_{\Delta\widetilde{t}}dt'\gamma_{\text{in}}\left(t'\right)}n_{\text{cr}}^{\text{L}}.
\end{equation}
where, once more, we have taken the limit $e^{-\Gamma_\text{L,R}}\rightarrow0$ only after integration. For the reversed pumping cycle, we merely have to replace the integral
$\int_{\Delta\widetilde{t}}dt'\rightarrow\int_{\Delta\widetilde{t}'}dt'$.

\section{Parameter independence of time interval differences.} \label{app_interval}
In Eq.~\eqref{eq_current_ratio} in the main text we argue that the time difference $\delta t_x$ is working point independent. In this Section we demonstrate this fact.

For a given pumping trajectory in energy space $\vec{\epsilon}(t)=(\epsilon_\text{L}(t),\epsilon_\text{R}(t))$, the kernel (describing the dynamics of the system) changes rapidly at some distinct points in time, namely at the crossing points between dot levels and between dot levels and Fermi energies of the reservoirs. As in the main text, we assume a sinusoidal time-dependence, $\epsilon_\text{L}(t)=\overline{\epsilon}_\text{L}+\delta\epsilon_\text{L} \sin(\Omega t)$ and $\epsilon_\text{R}(t)=\overline{\epsilon}_\text{R}+\delta\epsilon_\text{R} \sin(\Omega t+\varphi)$. We consider the values of $\varphi$ within the interval $[0,2\pi]$, while the amplitudes $\delta\epsilon_\text{L,R}$ are positive.
In the upcoming paragraph, we explicitly compute as a function of the driving parameters all the times where rapid changes in the dynamics of the DQD system occur. This enables us to show the working point independence of the current ratios.

When $\epsilon_\text{L}=\epsilon_\text{R}$, the system is for a very short period in the coherent regime. Such a level crossing occurs at,
\begin{eqnarray}
t_{\text{c}1}&=&\frac{\arcsin\left(\frac{\overline{\epsilon}_{\text{R}}-\overline{\epsilon}_{\text{L}}}{A}\right)-\delta}{\Omega}\\t_{\text{c}2}&=&\label{eq_tc2}\frac{-\pi-\arcsin\left(\frac{\overline{\epsilon}_{\text{R}}-\overline{\epsilon}_{\text{L}}}{A}\right)-\delta}{\Omega}\ ,
\end{eqnarray}
with
\begin{equation}
A=\sqrt{\delta\epsilon_{\text{L}}^{2}-2\delta\epsilon_{\text{L}}\delta\epsilon_{\text{R}}\cos\left(\varphi\right)+\delta\epsilon_{\text{R}}^{2}}\ ,
\end{equation}
and
\begin{equation}\label{eq_delta}
\begin{split}
\delta=\arctan\left(\frac{-\delta\epsilon_{\text{R}}\sin\left(\varphi\right)}{\delta\epsilon_{\text{L}}-\delta\epsilon_{\text{R}}\cos\left(\varphi\right)}\right)\\ -\pi\text{sign}\left[\sin(\varphi)\right]\theta\left[-\delta\epsilon_{\text{L}}+\delta\epsilon_{\text{R}}\cos\left(\varphi\right)\right]\ ,
\end{split}
\end{equation}
where the Heaviside $\theta$ function ensures that $\delta$ is a continuous function of $\varphi$ and $\delta\epsilon_\text{L,R}$, for $\varphi$ within the interval $[0,2\pi]$. A level crossing can only happen, when the condition  $|\overline{\epsilon}_{\text{L}}-\overline{\epsilon}_{\text{R}}|\leq |A|$ is fulfilled. Note that neither $A$ nor $\delta$ depend on the working point $(\overline{\epsilon}_\text{L},\overline{\epsilon}_\text{R})$.

Furthermore, when $\epsilon_\alpha=0$ ($\alpha=\text{L,R}$), the energy of dot $\alpha$ is in alignment with the chemical potential of electron reservoir $\alpha$. This event occurs at times,
\begin{eqnarray}
t_{\alpha1}&=&\frac{-\arcsin\left(\frac{\overline{\epsilon}_{\alpha}}{\delta\epsilon_{\alpha}}\right)-\varphi_{\alpha}}{\Omega}\\t_{\alpha2}&=&\frac{-\pi+\arcsin\left(\frac{\overline{\epsilon}_{\alpha}}{\delta\epsilon_{\alpha}}\right)-\varphi_{\alpha}}{\Omega}\ ,
\end{eqnarray}
with $\varphi_\text{L}=0$ and $\varphi_\text{R}=\varphi$.

We now show the working point independence of the current ratio for triangles A,B, and C.
For instance for triangle A, the relevant time difference appearing in the exponent of the current ratio, Eq.~\eqref{eq_ratio_GammaR}, is $\Delta t-\Delta t'$, where $\Delta t$ ($\Delta t'$) is the time difference between the event $\epsilon_\text{R}=0$ and the level crossing event $\epsilon_\text{L}=\epsilon_\text{R}$ for the forward (backward) trajectory. In order to relate $\Delta t-\Delta t'$ with the crossing times $t_{\text{c}1,2}$ and $t_{\alpha 1,2}$ the order of the different crossing events is important. For instance, suppose that we have a trajectory for which $t_{\text{R}2}<t_{\text{c}1}<t_{\text{c}2}<t_{\text{R}1}$. Then we find $\Delta t-\Delta t'=t_{\text{c}1}-t_{\text{R}2}+t_{\text{c}2}-t_{\text{R}1}$, that is,
\begin{equation}
\Delta t-\Delta t'=2\frac{\varphi-\delta}{\Omega}\ ,
\end{equation}
which is manifestly working point independent. Note however, that for a general trajectory characterized by $\delta\epsilon_\text{L,R}$ and $\varphi$, the order of the crossing times may be different. In this case, one has permute their order by shifting some of the crossing times by one period $\pm\tau$. In this case, we have to add $n\tau$ to above equation, where the integer $n$ accounts for such shifts.

For triangle C, the situation is similar, except that $\Delta\tilde{t}-\Delta\tilde{t}'$ depends on the time intervals between the events $\epsilon_\text{L}=0$ and $\epsilon_\text{L}=\epsilon_\text{R}$. Hence, 
\begin{equation}
\Delta\tilde{t}-\Delta\tilde{t}'=-2\frac{\delta}{\Omega}\ ,
\end{equation}
that is, with respect to the expression for $\Delta t-\Delta t'$, the phase shift $\varphi$ for the sinusoidal time-dependence of right dot disappears. 

Triangle B has not been explicitly addressed in this manuscript. We however, for completeness also show the working-point independence in this case. For triangle B, no level crossing takes place, and the relevant time intervals are $\Delta T$ ($\Delta T'$), namely the time interval between the events $\epsilon_\text{L}=0$ and $\epsilon_\text{R}=0$ for a certain forward (backward) trajectory. We now need to check the working point independence of the time difference $\Delta T-\Delta T'$. Suppose that the ordering of events is such that $t_{\text{L}1}<t_{\text{R}1}<t_{\text{R}2}<t_{\text{L}2}$, then
\begin{equation}
\Delta T-\Delta T'=-2\frac{\varphi}{\Omega}
\end{equation}
which is indeed also independent of the working point.

\bibliographystyle{apsrev4-1}
\bibliography{bib_phd_tot.bib}

\end{document}